\documentclass[aps,prd,showpacs,nofootinbib,floats,floatfix,preprintnumbers,groupedaddress,twocolumn]{revtex4-2}

\usepackage{aas_macros}

\usepackage[normalem]{ulem} 
\usepackage{bm}
\usepackage[titletoc]{appendix}
\usepackage{latexsym}
\usepackage{dcolumn}
\usepackage{amsmath,amsfonts,amssymb}
\usepackage{graphicx,epsfig}
\usepackage{amsmath}
\usepackage{fancyhdr}
\usepackage{hyperref}
\usepackage{graphicx,epstopdf}
\usepackage{xcolor}
\hypersetup{
	colorlinks   = true, 
	urlcolor     = blue, 
	linkcolor    = blue, 
	citecolor   = red 
}

\def\a{\alpha}

\def\bea{\begin{eqnarray}}
	\def\eea{\end{eqnarray}}

\begin{document}

\title{Study of relativistic hot accretion flow around Kerr-like Wormhole}
\author{Gargi Sen}\email{g.sen@iitg.ac.in}
\author{Debaprasad Maity}\email{debu@iitg.ernet.in}
\author{Santabrata Das}\email{sbdas@iitg.ac.in (Corresponding Author)}
	
\affiliation{Department of Physics, Indian Institute of Technology Guwahati, Guwahati 781039, Assam, India
	}
\date{\today}
	
\begin{abstract}

We investigate the structure of relativistic, low-angular momentum, inviscid advective accretion flow in a stationary axisymmetric Kerr-like wormhole (WH) spacetime, characterized by the spin parameter ($a_{\rm k}$), the dimensionless parameter ($\beta$), and the source mass ($M_{\rm WH}$). In doing so, we self-consistently solve the set of governing equations describing the relativistic accretion flow around a Kerr-like WH in the steady state, and for the first time, we obtain all possible classes of global accretion solutions for transonic as well as subsonic flows. We study the properties of dynamical and thermodynamical flow variables and examine how the nature of the accretion solutions alters due to the change of the model parameters, namely energy ($\mathcal{E}$), angular momentum ($\lambda$), $a_{\rm k}$, and $\beta$. Further, we separate the parameter space in $\lambda-\mathcal{E}$ plane according to the nature of the flow solutions, and study the modification of the parameter space by varying $a_{\rm k}$ and $\beta$. Moreover, we retrace the parameter space in $a_{\rm k}-\beta$ plane that allows accretion solutions containing multiple critical points. Finally, we calculate the disc luminosity ($L$) considering free-free emissions for transonic solutions as these solutions are astrophysically relevant and discuss the implication of this model formalism in the context of astrophysical applications.

\end{abstract}
	
\maketitle
	
\section{Introduction}

Accretion is believed to be the fundamental mechanism \cite{Frank-etal2002} that successfully explains the origin as well as the nature of the characteristic radiations that are emerged out from the astrophysical sources, namely quasars \cite{Smith-1966, Martin-2000}, active galactic nuclei \cite{Peterson-Bradley1997, Fabian-1999}, and black hole X-ray binaries \cite{Vidal-etal1973}. In the standard general relativistic framework, a massive compact object at the center of the accreting system plays a central role in this accretion process. Out of different theoretical possibilities as central objects, black hole (BH) makes the phenomena extremely interesting because of its unique underlying characteristics at the event horizon. Hence, many theoretical studies on accretion process have been confined to those systems, where BH assumes the role of central object \cite[]{Pringle-Rees1972,Shakura-Sunyaev1973,Novikov-Thorne1973}. However, from the observational point of view, it is not the BH that can be observed directly. Therefore, in the theoretical front, the candidate for the gravitating central object can be any consistent solution of general relativity that seems to mimic the black hole space-time in the asymptotic region. Accordingly, in this endeavour, the primary motivation would be to study the properties of the accreting system in the strong gravity regime, which is yet to be proved strictly to be Einsteinian. 

Meanwhile, recent observations of black hole shadows by the Event Horizon Telescope (EHT) \cite{Akiyama-etal2019a, Akiyama-etal2019, Akiyama-etal2022a, Akiyama-etal2022} have opened up the possibility of detecting the direct signature of strong gravity. In this domain, there has been a significant surge for the investigation of various exotic gravitational objects in recent years. Specifically, an exotic gravitational background obtained within the framework of general relativity could be intriguing to explore in the context of the strong gravity regime through accretion processes. It is worth mentioning that BHs need not be the only accreting objects in the universe, instead there maybe other categories of hypothetical objects, such as naked singularity (NS) and wormhole (WH), which can not be ruled out by theory and/or experiment till date. Indeed, WHs are the valid solutions of Einstein’s equations similar to BHs, and hence, it has been an active area of research to study the accretion phenomenon around them. Meanwhile, numerous attempts along this line were carried out adopting different gravitational theories, such as higher dimensional braneworld gravity\cite{Pun-etal2008, Heydari-2010}, Chern-Simons modified gravity \cite{Harko-etal2010}, Ho{\v{r}}ava gravity \cite{lu-2009}, more exotic boson stars \cite{TORRES-2002, Guzman-2006}, wormholes \cite{Harko-etal2009a}, gravastars \cite{Harko-etal2009}, quark stars \cite{Kovacs-2009}. Needless to mention that all these works were performed considering an incomplete description of the accretion flow, particularly taking into account only the particle dynamics \cite {Shakura-Sunyaev1973, Novikov-Thorne1973, Page-Thorne1974, Thorne-1974}. Recently, a full general relativistic hydrodynamic treatment is reported \cite{Dihingia-etal2020,Sen-etal2022} for a special class of background called Kerr-Taub-NUT (KTN) spacetime and a complete set of accretion solutions, and their properties are discussed.
	
Keeping this in mind, in the present paper, we take up another class of spacetime called Kerr-like WH which has recently gained widespread interest in the astrophysical context \cite[]{Bueno-etal2018}. After the very first proposal of Einstein and Rosen \cite{Einstein-Rosen1935} with unsuccessfully countering the non-local nature of quantum mechanics, famously known as Einstein–Rosen bridge, significant efforts have been imparted over the years to understand such exotic object in purely Einstein's framework \cite{Wheeler-1955, Fuller-Wheeler1962}, adding minimally coupled scalar field with negative kinetic term \cite{Ellis-1973,Bronnikov-1973}. In general relativity framework, exotic matter violating null, weak, and strong energy conditions \cite{Morris-Thorne1988, Morris-etal1988, Visser-etal2003, kar-etal2004, Lobo-2007, Visser-1989} has been observed to play crucial role in generating WH solution. These include models which are supported by the phantom energy, the cosmological constant \cite{ Cataldo-etal2009, Lemos-etal2003, Rahaman-etal2007, Lobo-2005b}, modified theories of gravity such as higher order curvature theory \cite{Harko-etal2013}, non-minimal curvature-matter coupling in a generalized $f(R)$ modified theory of gravity \cite{Garcia-Lobo2011}, modified theories, $e. g.$, Einstein–Gauss–Bonnet \cite{Mehdizadeh-etal2015}, Born–Infeld gravity \cite{Shaikh-2018}, Einstein–Cartan \cite{Mehdizadeh-Ziaie2017}. WHs generically come with a throat that connects two different asymptotic regions, and away from the throat, the WHs mimics black hole spacetime \cite{Cardoso-Pani2019}. There have been several works on the possibility to distinguish the classical WHs from BHs by means of various diagnostics, such as the shadow of an accretion disc, gravitational lensing, gravitational waves, etc., \cite{Harko-etal2009a, Konoplya-Zhidenko2016, Shaikh-Kar2017, Shaikh-2018, Amir-etal2019, Karimov-etal2019,  Kasuya-Kobayashi2021, Stuchlik-Vrba2021, Jusufi-etal2022, Kiczek-Rogatko2022}. However, a complete hydrodynamical analysis of accretion process are still pending in WH background. This motivates us to investigate the hydrodynamic properties of accretion flow around Kerr-like WH in full general relativistic framework.
 
This paper is organized as follows. In \S II, we describe the background geometry. In \S III, we present the underlying assumptions and governing equations. We present the method to find the global transonic and subsonic solutions around WH in \S IV. In \S V, we present the obtained results. We discuss the radiative emission properties in \S VI. Finally, we summarize our findings with conclusions in \S V.

\section{Background Geometry}
	
We begin with a stationary, axisymmetric, Kerr-like WH spacetime \cite{Bueno-etal2018}, where the spacetime interval is expressed as,
\begin{equation}
    \begin{aligned}
    ds^2 = & g_{\mu\nu}dx^\mu dx^\nu\\
    =& g_{tt}dt^2 + g_{rr}dr^2 + 2 g_{t\phi} dt d\phi + g_{\phi\phi} d\phi^2 +g_{\theta\theta} d\theta^2.
    \label{metric}
    \end{aligned}
\end{equation}
Here, the coordinate $r$ globally defines the WH spacetime in the following ways. We impose a discrete $Z_2$ symmetry on $r$ such that, $0 \le |r| \le \infty$. In both cases, the wormhole spacetime is globally static, and the time-like Killing vector remains time-like everywhere. On the contrary, in the black hole scenario, time-like spacetime becomes space-like beyond the horizon.
	
\begin{figure}
    \begin{center}
    \includegraphics[width=0.8\columnwidth]{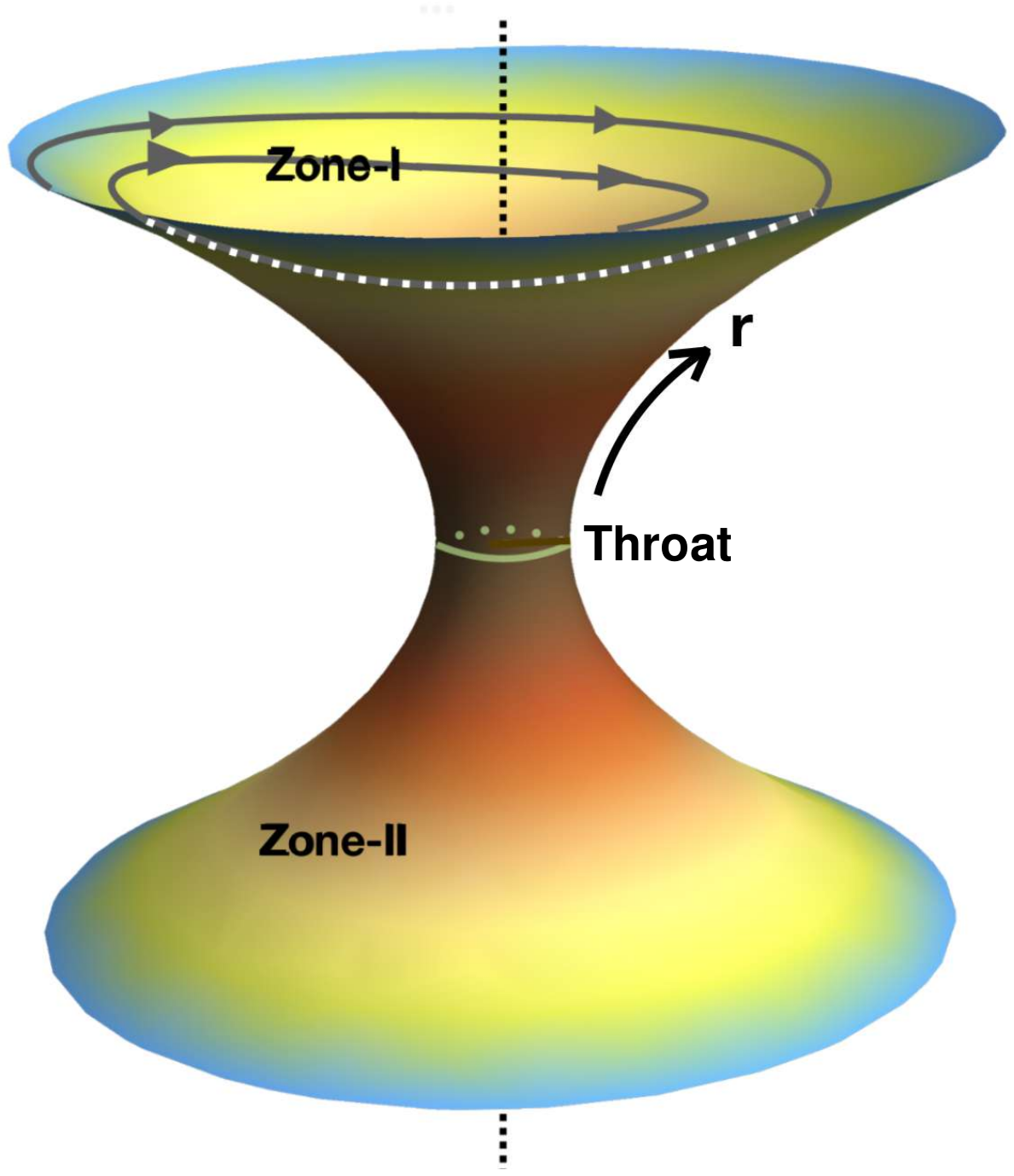}
    \end{center}	
    \caption{Artistic impression of a symmetric WH spacetime that includes Zone-I and Zone-II connected via throat.
    }
    \label{WH-space}	
\end{figure}

Considering the symmetric WH, the metric components in both sides of the WH throat are obtained in terms of Boyer-Lindquist coordinates \cite{Boyer-etal1967}, which are given by, 
\begin{align*}
    g_{tt}|_{\pm}&= -(1-\frac{2r}{\Sigma})~~;~~ ~g_{t\phi}|_{\pm}=-\frac{2a_{\rm k}r\sin^{2}\theta}{\Sigma}\;;\\
    g_{rr}|_{\pm}&=\frac{\Sigma}{\Delta}~;~   
    g_{\theta\theta}|_{\pm}=\Sigma ;\\
    g_{\phi\phi}|_{\pm}&= (r^2 + a_{\rm k}^2 + \frac{2 a_{\rm k}^2 r \sin^2\theta}{\Sigma})\sin^2\theta,
\end{align*}
where, $\Sigma= r^{2}+a^{2}_{\rm k}\cos^{2}\theta$, $\Delta=r^{2}-2r (1+\beta^2)+a^{2}_{\rm k}$, $a_{\rm k}$ is the spin parameter (equivalently Kerr parameter), and $\beta$ is the dimensionless parameter. Here, `$\pm$' denotes two sides of the WH under consideration, and we refer `$+$' to Zone-I and `$-$' to Zone-II as illustrated in Fig. \ref{WH-space}. For a limiting value $\beta =0$, the Kerr-like WH turns out to be a Kerr black hole.

In these analysis, we follow the sign convention as ($-, +, +, +$) and adopt a unit system $M_{\textrm{WH}} = G = c = 1$, where $M_{\textrm{WH}}$ denotes the WH mass, $G$ is the universal gravitational constant, and $c$ is the speed of light. In this unit system, length, time, and angular momentum are expressed in units of $GM_{\rm WH}/c^2$, $GM_{\rm WH}/c^3$, and $GM_{\rm WH}/c$, respectively. The property of a stationary axisymmetric spacetime is the existence of two commuting killing vectors along $(t,\phi)$ directions. The other two components $(r,\theta)$ are mutually orthogonal to each other. Setting the condition $g^{rr}=1/g_{rr}=0$, we calculate the throat radius as $r_{\rm th}= (1+\beta^2)^2 + \sqrt{(1+\beta^2)^2-a^2_{\rm k}}$. In this work, we consider traversable WH, where, depending on the appropriate boundary conditions, accreting matter from one Zone can smoothly pass to the other Zone via throat. Hence, in order to study the properties of accretion flow, we analyze the governing equations for both sides (Zone-I and Zone-II) of the throat.

\section{Assumptions and governing equations}

We consider a low angular momentum, steady, inviscid, axisymmetric, advective accretion flow around a WH. In addition, the flow is assumed to remain confined at the equatorial plane of the central object and flow does not suffer energy dissipation due to various physical processes, namely viscosity, radiative cooling and magnetic fields.

In the general relativistic hydrodynamic framework, the energy-momentum tensor and four current are given by,
\begin{equation}
    T^{\mu\nu} = (e + p)u^{\mu}u^{\nu} + pg^{\mu\nu} \quad \text{and} \quad j^{\mu} = \rho u^{\mu},
    \label{emtensor}
\end{equation}
where $e$, $p$, $\rho$, and $u^\mu$ denote the internal energy density, pressure, mass density and four velocities of the perfect fluid, respectively and the spacetime indices $\mu$ and $\nu$ run from $0$ to $3$.
	
The hydrodynamical accretion flow is governed by conservation of energy-momentum and mass flux equations, which are given by,
\begin{equation}
    T^{\mu\nu}_{;\nu} = 0 \quad\text{and} \quad (\rho u^{\nu})_{;\nu} = 0.
    \label{coneqn}
\end{equation}
Here, the time-like velocity field obeys the condition $u_\mu u^\mu =-1$. We use the projection operator defined as $h^{i}_{\mu}=\delta^i_{\mu}+u^iu^{\mu}$ to take the projection of the conservation equation on the spatial hypersurface and obtain the Euler equation as,
\begin{equation}
    h^{\alpha}_{\mu} T^{\mu\nu}_{;\nu} = (e+p)u^{\nu}u^{\alpha}_{;\nu} + (g^{\alpha \nu}+u^{\alpha}u^{\nu})p_{,\nu}=0.
    \label{euler}
\end{equation}
Note that the projection operator also satisfies the condition $h^{\alpha}_{\mu} u^\mu =0$ which ensures that the projection operator and the four velocity remain orthogonal to each other. Further, we project the conservation equation along $u^\mu$ and obtain the first law of thermodynamics as, 
\begin{equation}
    u_{\mu} T^{\mu\nu}_{;\nu}= u^{\nu}\left[\left(\frac{e + p}{\rho}\right)\rho_{,\nu}-e_{,\nu}\right]=0.
    \label{entropygen}
 \end{equation}
In this work, we assume the flow to remain confined around the disk equatorial plane and hence, we choose $\theta=\pi/2$ which leads to $u^\theta=0$. Further, following \cite{Lu-1985}, we define the three radial velocity of the fluid in the co-rotating frame as $v^{2}=\gamma^{2}_{\phi}v^{2}_{r}$, where $\gamma^{2}_{\phi}=1/(1-v^{2}_{\phi})$, $v^{2}_{\phi}=(u^{\phi}u_{\phi})/(-u^{t}u_{t})$, and $v^{2}_{r}=(u^{r}u_{r})/(-u^{t}u_{t})$, respectively. The radial Lorentz factor $\gamma^{2}_{v} = 1/(1-v^{2})$ and the total bulk Lorentz factor is $\gamma = \gamma_{\phi}\gamma_{v}\gamma_{\theta}$. With the above definitions of velocities, we obtain the radial component of the momentum equation from Eq. (\ref{euler}) for $\alpha = r$ as,
\begin{equation}
    v\gamma_{v}^{2}\frac{dv}{dr}+\frac{1}{h\rho}\frac{dp}{dr}+\frac{d\Phi^{\textrm{eff}}_{\rm e}}{dr}=0,
    \label{eulerradial}
\end{equation}
where $h\left[ =(e+p)/\rho\right]$ is the specific enthalpy, $\Phi^{\rm eff}_{\rm e}$ refers the effective potential \cite{Dihingia-etal2018} at the disk equatorial plane and is given by,
\begin{equation}
    \Phi^{\textrm{eff}}_{\rm e}=1+\frac{1}{2}\ln\left[\frac{r( a_{\rm k}^2 +r(r-2)) }{a_{\rm k}^{2}(r+2)-4a_{\rm k}\lambda+r^{3}-\lambda^{2}(r-2)}\right].
    \label{potential}
\end{equation}

 \begin{figure}
    \begin{center}
    \includegraphics[width=\columnwidth]{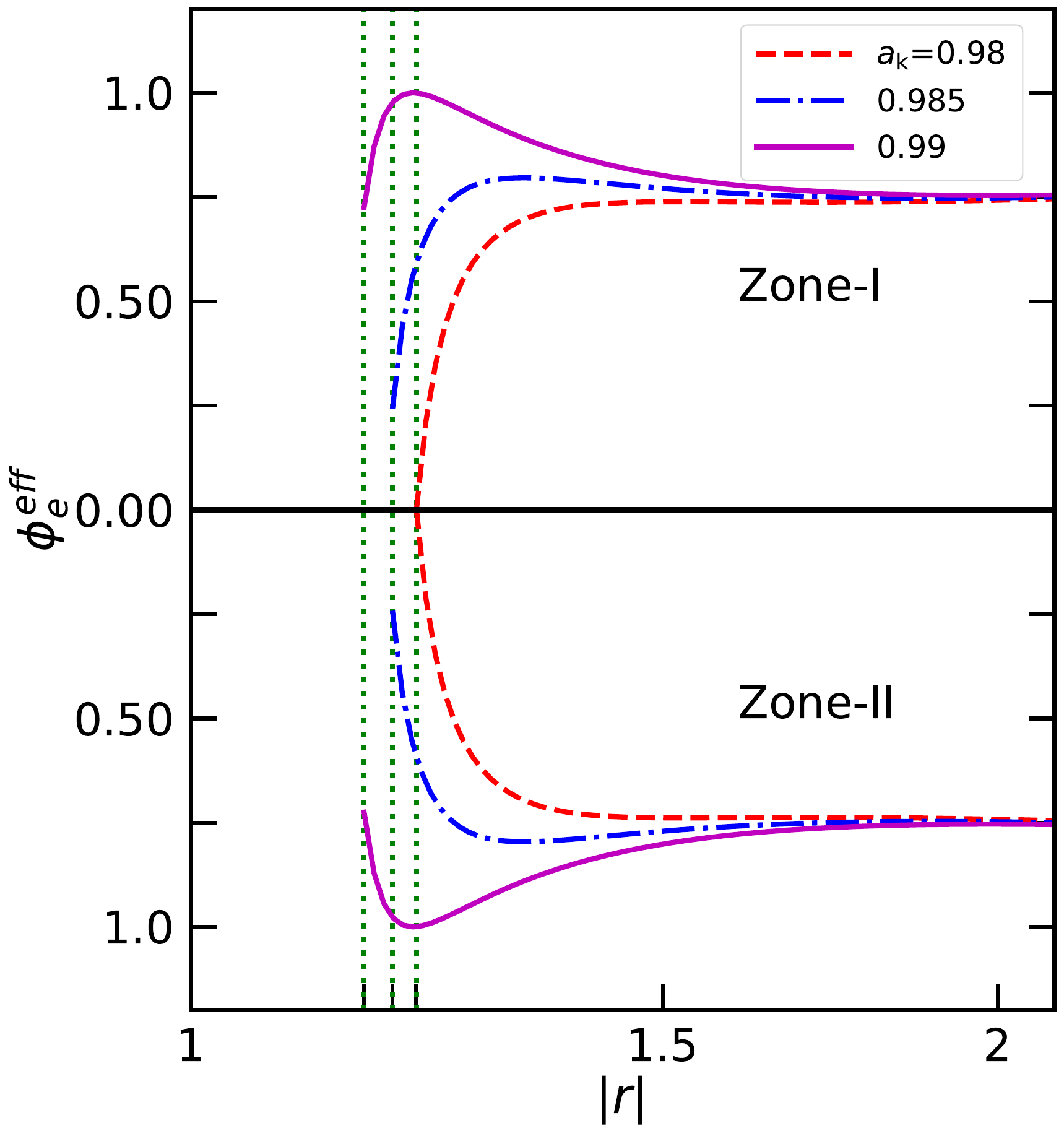}
    \end{center}
    \caption{Variation of effective potential ($\Phi^{\textrm{eff}}_{\rm e}$) as function radial coordinate ($|r|$; modulus is used for the simultaneous representation of Zone-I and Zone-II) for angular momentum $\lambda = 2.20$. Dashed (red), dot-dashed (blue) and solid (magenta) curves denote results corresponding to $a_{\rm k}=0.98$, $0.985$, and $0.99$, respectively, and dotted (green) vertical lines indicate the respective throat radius as $r_{\rm th}=1.2137$, $1.1890$, and $1.1603$. See the text for the details.
    }
    \label{poten_3ak_3beta}
\end{figure}

Needless to mention that the overall characteristics of the accretion flow crucially depend on the nature of the gravitational potential outside WH under consideration. Hence, we examine the effective potential ($\Phi^{\rm eff}_{\rm e}$) in Fig. \ref{poten_3ak_3beta}, where the variation of $\Phi^{\rm eff}_{\rm e}$ with radial coordinate ($r$) is illustrated for a fixed angular momentum $\lambda = 2.20$. In the figure, the obtained results are plotted with dashed (red), dot-dotted (blue) and solid (magenta) curves for $a_{\rm k} = 0.98$, $0.985$, and $0.99$, respectively. The dotted (green) vertical lines denote the throat radius ($r_{\rm th}$) of WH that solely depends on both $a_{\rm k}$ and $\beta$, respectively. Here, we choose $\beta = 0.05$ and find $r_{\rm th} = 1.2137$, $1.1890$, and $1.1603$ for the chosen spin parameters ($a_{\rm k}$) in increasing order. Note that the horizontal solid (back) line separates Zone-I from Zone-II on both sides of the WH throat. Figure evidently indicates that the potential is symmetric in both sides (Zone-I and Zone-II) of WH throat. 

Using Eq. (\ref{entropygen}), we obtain the entropy generation equation along the radial direction as,
\begin{equation}
    \left( \frac{e+p}{\rho}\right)\frac{d\rho}{dr}-\frac{de}{dr}=0.
    \label{entgen}
\end{equation}

The stationary and axisymmetric spacetime under consideration is associated with two Killing vectors due to its symmetries. This yields two conserved quantities, which are given by, 
\begin{equation}
    -h u_{t} = {\cal E}  \quad; \quad h u_{\phi} = {\cal L},
    \label{energy-angularmomentum}
\end{equation}
where, $\cal E$ is the Bernoulli constant (equivalently specific energy) and $\cal L$ is the bulk angular momentum per unit mass of the flow. We express the specific angular momentum of the flow as $\lambda = {\cal L}/{\cal E}=-u_{\phi}/u_{t}$, which is also a conserved quantity for an inviscid accretion flow.

We integrate Eq. (\ref{coneqn}) to obtain another constant of motion in the form of mass accretion rate ($\dot M$) and is given by,
\begin{equation}
    \dot{M}=-4\pi r \rho u^{r} H.
    \label{mdot}
\end{equation}
In this work, we express the mass accretion rate in dimensional form as ${\dot m}={\dot M}/{\dot M}_{\rm Edd}$, where ${\dot M}_{\rm Edd} ~(=1.44 \times 10^{18}(M_{\rm WH}/M_\odot)$ gm s$^{-1}$) is the Eddington accretion rate, $M_\odot$ being the solar mass. In Eq. (\ref{mdot}), $H$ refers the local half-thickness of the accretion disk. Following \cite{Riffert-Herold1995, Peitz_Appl1997}, we compute $H$ assuming the flow to maintain hydrostatic equilibrium in the vertical direction, and is given by
\begin{equation}
    H=\sqrt{\frac{pr^{3}}{\rho F}}; \quad F=\frac{1}{1-\lambda\Omega} \times \frac{(r^{2}+a_{\rm k}^{2})^{2}+2\Delta a_{\rm k}^{2}}{(r^{2}+a_{\rm k}^{2})^{2}-2\Delta a_{\rm k}^{2}},
    \label{height}
\end{equation}
where $\Omega ~ [=(2a_{\rm k}+\lambda(r-2))/(a^2_{\rm k}(r+2)-2a_{\rm k}\lambda+r^3)]$ is the angular velocity of the accreting matter.

We close the equations (\ref{coneqn}) and (\ref{mdot}) adopting the relativistic equation of state (REoS) \cite{Chattopadhyay-Ryu2009} that relates internal ($e$), pressure ($p$) and mass density ($\rho$) as,
\begin{equation}
    e = \frac{\rho f}{\tau}, \quad p=\frac{2\rho\Theta}{\tau},
    \label{eos}
\end{equation}
with $\tau = 1+ m_p/m_e$ and 
$$
f=\left[ 1+ \Theta \left( \frac{9\Theta +3}{3 \Theta +2}\right) \right] + \left[ \frac{m_p}{m_e} + \Theta \left( \frac{9\Theta m_e +3m_p}{3 \Theta m_e+ 2 m_p}\right) \right],
$$
where $m_p$ and $m_e$ denote the masses of ion and electron, respectively, and $\Theta~(=k_{\rm B}T/m_e c^2)$ is the dimensionless temperature. In accordance with REoS, the speed of sound is expressed as $C_s = \sqrt{2\Gamma\Theta/(f+2\Theta)}$, where $\Gamma ~ [= (1+N)/N]$ refers the adiabatic index and $N ~[=(1/2)(df/d\Theta)]$ is the polytropic index of the flow, respectively \cite{Dihingia-etal2019}. Using Eq. (\ref{entgen}), we estimate the measure of entropy by calculating the entropy accretion rate ($\dot{\mathcal{M}}$) \cite{Chattopadhyay-Ryu2009, Dihingia-etal2020}, which is given by,
\begin{equation}
    \dot{\mathcal{M}}=\exp(k_1) \Theta^{3/2} \left(2 + 3 \Theta\right)^{3/4} \left(3 \Theta + \frac{2m_p}{m_e}\right)^{3/4} u^r r H,
    \label{entacc}
\end{equation}
where $k_1 = \left[ f - \left(1+m_p/m_e\right)\right]/2\Theta$. Note that for a non-dissipative flow characterized with a given set of energy ($\mathcal{E}$) and angular momentum ($\lambda$), $\dot{\mathcal{M}}$ remains conserved all throughout the disk.

We simplify Eqs. (\ref{eulerradial}), (\ref{entgen}), (\ref{energy-angularmomentum}), (\ref{mdot}) and (\ref{eos}) and obtain the wind equation as,
\begin{equation}
    \frac{dv}{dr}=\frac{\mathcal{N}}{\mathcal{D}},
    \label{dvdr}
\end{equation}
where the numerator ($\mathcal{N}$) is given by,
\begin{equation}
    \mathcal{N}=\frac{2 C^{2}_{s}}{\Gamma +1}\left(\frac{1}{2{\Delta}}\frac{d{\Delta}}{dr}+\frac{3}{2r}-\frac{1}{2F}\frac{dF}{dr}\right)-\frac{d\Phi^{\text{eff}}_{\rm e}}{dr},
    \label{numerator}
\end{equation}
and the denominator ($\mathcal{D}$) is given by, 
\begin{equation}
    \mathcal{D}=\gamma_{v}^{2}\left(v-\frac{2 C^{2}_{s}}{v(\Gamma +1)}\right).
    \label{denominator}
\end{equation}
Further, using Eqs. (\ref{mdot}), (\ref{height}) and (\ref{dvdr}), we calculate the radial gradient of the dimensionless temperature as,
\begin{equation}
    \frac{d\Theta}{dr}=\frac{-2\Theta}{2N+1}\left[\frac{1}{2{\Delta}}\frac{d{\Delta}}{dr}+\frac{3}{2r}+\frac{\gamma_{v}^{2}}{v}\frac{dv}{dr}-\frac{1}{2F}\frac{dF}{dr}\right].
    \label{dthetadr}
\end{equation}

\section{Solution Methodology}

During the course of accretion around WH, rotating flow from the outer edge ($r_{\rm edge}$) of the disk in Zone-I (Zone-II) starts accreting subsonically ($v<C_s$). Because of the strong gravity of WH, inward moving flow gradually gains its radial velocity and depending of the input parameters, namely $\cal E$, $\lambda$, $a_{\rm k}$ and $\beta$, flow may become super-sonic after crossing the critical point ($r_{\rm c}$; flow of this kind is called transonic flow) or remain subsonic all throughout before approaching to the WH throat ($r_{\rm th}$). Thereafter, flow is diverted to Zone-II (Zone-I) with identical velocity ($v$), temperature ($\Theta$) and accretion rate ($\dot M$) at $r_{\rm th}$ of Zone-I (Zone-II), and continues to proceed away from the WH till $r_{\rm edge}$. It is noteworthy that a transonic (subsonic) flow in Zone-I remains transonic (subsonic) in Zone-II, and vice versa. 

\subsection{Transonic Accretion Solutions}

In general, the accretion flow around WH remains smooth everywhere ($r_{\rm th} < r \le r_{\rm edge}$), and hence, the flow radial velocity gradient $(dv/dr)$ must be real and finite along the flow streamline. However, equation (\ref{denominator}) clearly indicates that the denominator ($\mathcal{D}$) may vanish at some points. If so, numerator ($\mathcal{N}$) also vanishes there. Such a special point, where $\mathcal{N}=\mathcal{D}=0$, is called as critical points ($r_{\rm c}$). Setting the condition $\mathcal{D}=0$, we obtain the radial velocity ($v_{\rm c}$) of the flow at the critical point ($r_{\rm c}$) as,
\begin{equation}
    v_{\rm c}=\sqrt{\frac{2 }{\Gamma_{\rm c} +1} }C_{\rm sc}.
    \label{vc}
\end{equation}
Similarly, the condition $\mathcal{N}=0$ yields the sound speed ($C_{\rm sc}$) at $r_{\rm c}$ as,
\begin{equation}
    C^{2}_{\rm sc}=\frac{\Gamma_{\rm c} +1}{4}\left(\frac{d\Phi^{\textrm{eff}}_{\rm e}}{dr}\right)_{\rm c}
    \left(\frac{1}{2{\Delta}}\frac{d{\Delta}}{dr}+\frac{3}{2r}-\frac{1}{2F}\frac{dF}{dr}\right)_{\rm c}^{-1}.
    \label{csc}
\end{equation}
In equation (\ref{vc}) and (\ref{csc}), quantities with subscript `c' are evaluated at the critical point ($r_{\rm c}$).

As the radial velocity gradient $(dv/dr)$ takes $0/0$ form at $r_{c}$, we apply L$'$H\^{o}pital's rule to evaluate $(dv/dr)_{\rm c}$ at $r_{\rm c}$. For a given set of input parameters ($\cal E$, $\lambda$, $a_{\rm k}$ and $\beta$), $(dv/dr)_{\rm c}$ yields two values. When both $(dv/dr)_{\rm c}$ are real and of opposite sign, the critical point is called as saddle type, whereas nodal type critical point is obtained if $(dv/dr)_{\rm c}$ are real and of the same sign. For spiral type critical point, $(dv/dr)_{\rm c}$ are imaginary. Needless to mention that saddle type critical points are stable, whereas both nodal and spiral types critical points are unstable \cite{Kato-etal1993}. Hence, saddle type critical points are specially relevant in the astrophysical context as transonic accretion solution can only pass through them. Now onwards, we refer saddle type critical points as critical points only unless stated otherwise. Furthermore, depending on the input parameters, flow may possess more than one critical points. When critical point forms close to throat, it is called as inner critical point $(r_{\textrm{in}})$ and when it forms far away from the throat is referred as outer critical point $(r_{\textrm{out}})$.

In order to obtain the self-consistent transonic solution around WH, we simultaneously solve  Eq. (\ref{dvdr}) and Eq. (\ref{dthetadr}) for a given set of input parameters ($\cal E$, $\lambda$, $a_{\rm k}$, $\beta$) in Zone-I (Zone-II). In doing so, we first integrate Eq. (\ref{dvdr}) and Eq. (\ref{dthetadr}) starting from the critical point ($r_{\rm c}$) up to the outer edge of the disk ($r_{\rm edge}$) and then from $r_{\rm c}$ to $r_{\rm th}$. Finally, we join these two segments of the solution to obtain the global transonic solution in Zone-I (Zone-II). It is worth mentioning that for a traversable WH, an accretion solution in Zone-I appears to be analogous in Zone-II.

\subsection{Subsonic Accretion Solutions}

Unlike transonic solution, subsonic solution does not pass through the critical point and hence, to obtain such solution uniquely, we require entropy accretion rate (${\dot {\cal M}}$) as additional parameter along with the other input parameters. Therefore, for a set of input parameters ($\cal E$, $\lambda$, $a_{\rm k}$, $\beta$, and ${\dot {\cal M}}$), we integrate Eq. (\ref{dvdr}) and Eq. (\ref{dthetadr}) starting from the outer edge of the disk ($r_{\rm edge}$) up to $r_{\rm th}$. To start the integration, we tune the flow radial velocity ($v_{\rm edge}$) at $r_{\rm edge}$ to calculate $\Theta_{\rm edge}$ using equation (\ref{entacc}), that renders smooth subsonic solution in the range $r_{\rm th} \lesssim r \le r_{\rm edge}$ in Zone-I (Zone-II). Note that for a given set of ($\cal E$, $\lambda$, $a_{\rm k}$, $\beta$), one can obtain a set of subsonic solutions around WH for different ${\dot {\cal M}}$ values.
	
\section{Results}

In this section, we present the results obtained form our model formalism that include the global solutions, parameter space and the emission properties of the accretion flow around WH. 

\begin{figure}
    \begin{center}
    \includegraphics[width=0.9\columnwidth]{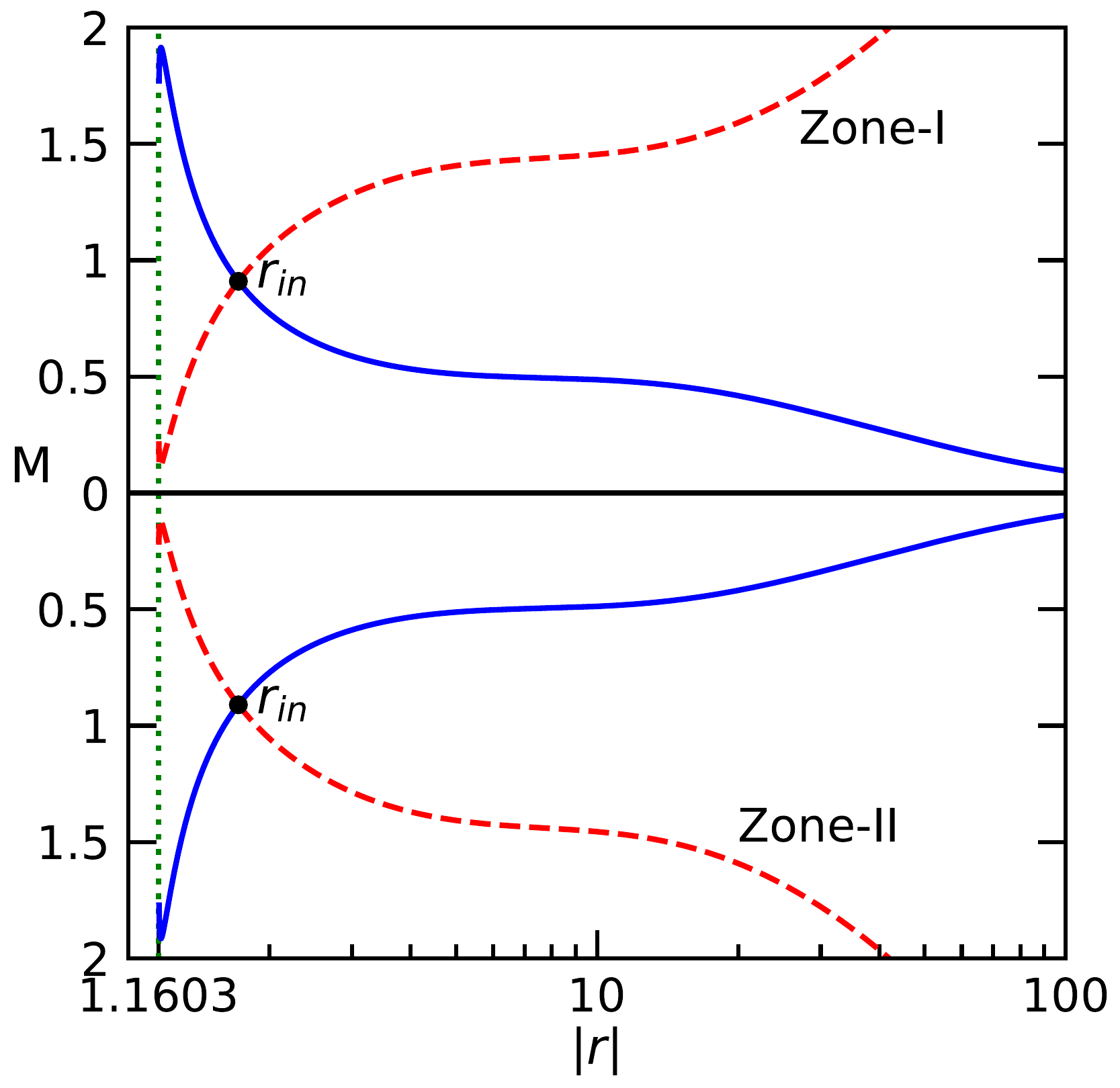}
    \end{center}
    \caption{Variation of Mach no $(M=v/C_{s})$ as function of the modulus of radial coordinate ($|r|$) around WH. Here, we choose $\mathcal{E}=1.02$, $\lambda=1.90$, $a_{\rm k} = 0.99$, and $\beta = 0.05$, respectively. Solid (blue) and dashed (red) curves represent solutions corresponding to accretion and winds. Filled circles (black) refer to the inner critical points ($r_{\rm in}$) and dotted vertical line (green) denotes throat radius of WH. See the text for the details.}
    \label{sol_i}
\end{figure}

\subsection{Global Transonic Solutions}

In Fig. \ref{sol_i}, we present a typical global transonic accretion solution where Mach number ($M=v/C_s$) of the flow is plotted as function of the radial coordinate ($r$). Here, we choose $a_{\rm k}$ = 0.99, and $\beta$ = 0.05, and the solutions are computed for flows of energy ${\cal E} = 1.02$ and angular momentum $\lambda=1.90$. For the chosen set of input parameters, we find that in Zone-I (upper panel), flow starts accreting subsonically from the outer edge of the disk at $r_{\rm edge}=100$ and gains radial velocity as it moves inward due the strong attraction of WH gravity. Eventually, flow changes its sonic state to become supersonic at the inner critical point at $r_{\rm in}=1.7160$ and continues to accrete until it reaches the WH throat at $r_{\rm th}=1.1603$. In the figure, we present the accretion solution using the solid (blue) curve. The corresponding wind solution (from $r_{\rm th}$ to $r_{\rm edge}$) is also depicted as shown by the dashed (red) curve. For the purpose of completeness, we present the flow solutions for Zone-II in the lower panel which is the mirror image of the flow solutions presented in the Zone-I. Now onwards, to avoid repetitions, we shall exclusively present the flow solutions in Zone-I only, unless stated otherwise.

\begin{figure}
    \begin{center}
    \includegraphics[width=0.9\columnwidth]{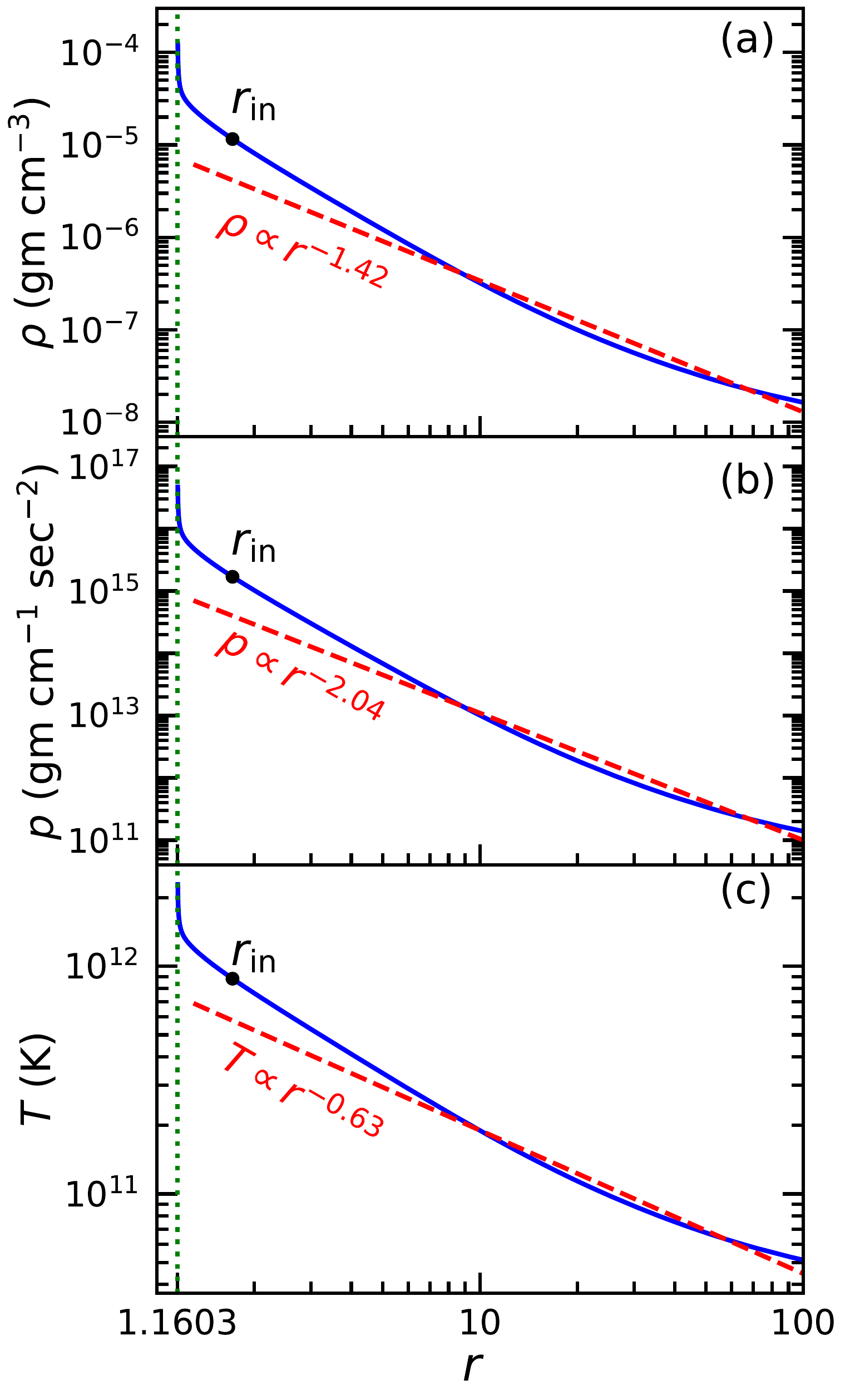}
    \end{center}
    \caption{Variation of (a) density ($\rho$), (b) pressure ($p$), and (c) temperature ($T$) of the accretion flow as function of radial coordinate ($r$). Here. model parameters are chosen same as in Fig. \ref{sol_i}. In each panel, dashed (red) curve represents the best fit power-law profile of the flow variables and filled circle (black) denotes the critical point $r_{\rm in}=1.7160$. The vertical dotted lines (green) denote throat radius $r_{\rm th} = 1.1603$. See the text for the details.}
    \label{sol_prop}
\end{figure}

 In Fig. \ref{sol_prop}, we display the variation of other flow variables with the radial coordinate ($r$) corresponding to the accretion solution presented in Fig. \ref{sol_i}. In Fig. \ref{sol_prop}a, we depict the profile of density ($\rho$) variation for convergent accretion flow and observe that $\rho$ increases as the flow moves towards WH. We put effort to represent the density profile using a power-law and the best fit is obtained as $\rho \propto r^{-(n+2/5)}$, where $n \sim 1$. This finding is consistent with the results reported in \cite[]{Narayan-Yi1995,Frank-etal2002}. Next, we show the radial profile of pressure ($p$) and temperature ($T$) of the flow in Fig. \ref{sol_prop}b-c, and attain the optimal power-law fit as $p \propto r^{-(n+1)}$ and $T \propto r^{-(n-1/3)}$, respectively. It is noteworthy that we observe poor fitting of the flow variables at the inner part of the disk close to WH. This possibly happens due to that fact that the simple power-law fit fails to capture the complex nature of the flow characteristics in the vicinity of the WH.
  
\begin{figure}
    \begin{center}
    \includegraphics[width=\columnwidth]{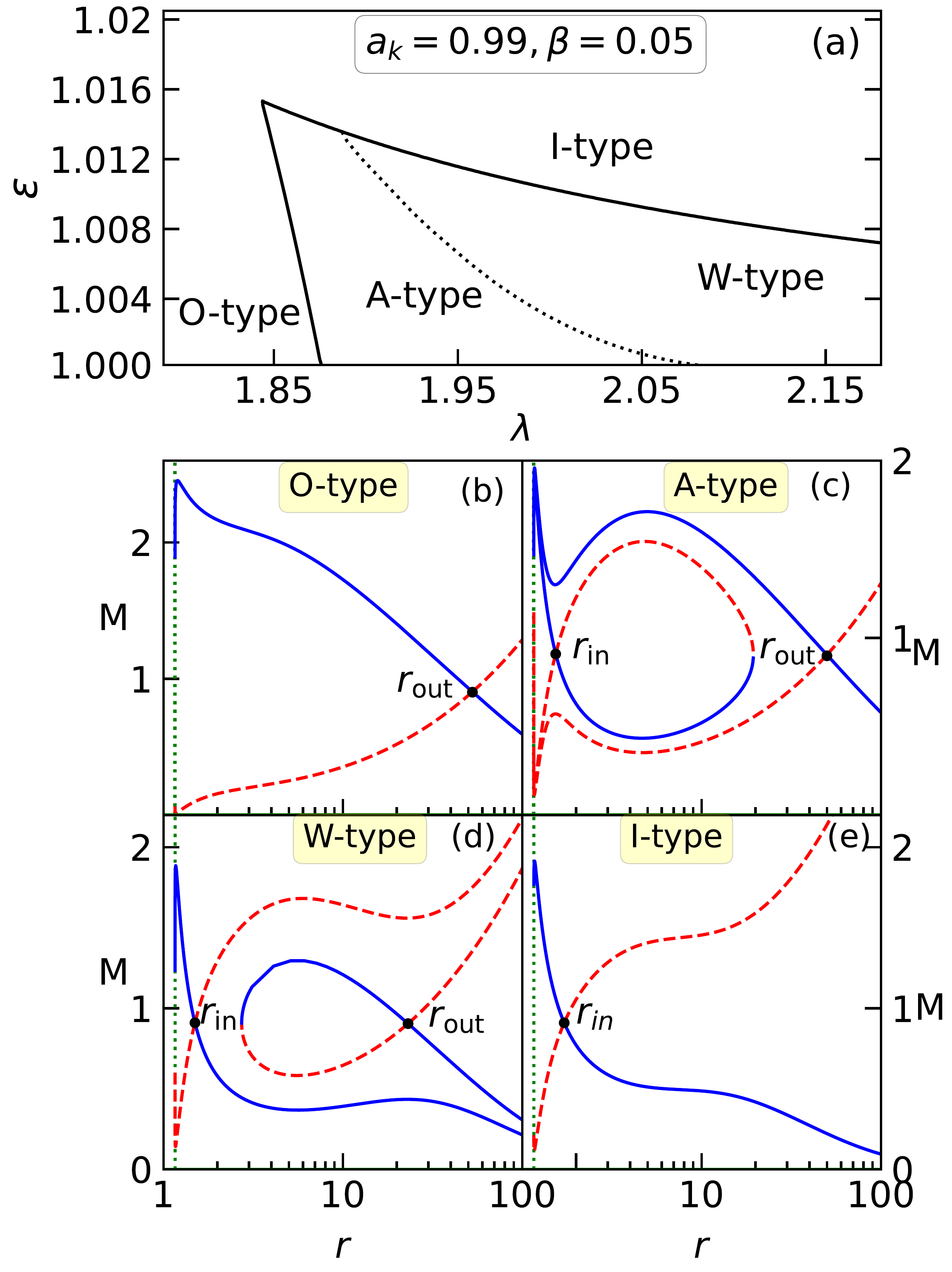}
    \end{center}
    \caption{Sub-division of parameter space in $\lambda-{\cal E}$ plane according to the nature of the flow solutions around WH (panel a). Here, we choose $a_{\rm k} = 0.99$ and $\beta = 0.05$. Four distinct regions marked as O-type, A-type, W-type and I-type are identified and typical flow solutions ($M$ vs. $r$) from these regions are depicted in panels (b-e), where solid (blue) and dashed (red) curves denote accretion and winds. Filled circles (black) refer critical points ($r_{\rm in}$ and/or $r_{\rm out}$) and vertical dotted lines (green) denote throat radius $r_{\rm th} = 1.1603$. See the text for the details.
 	}
    \label{para-sol}
\end{figure}

\begin{figure*}
    \begin{center}
    \includegraphics[width=\textwidth]{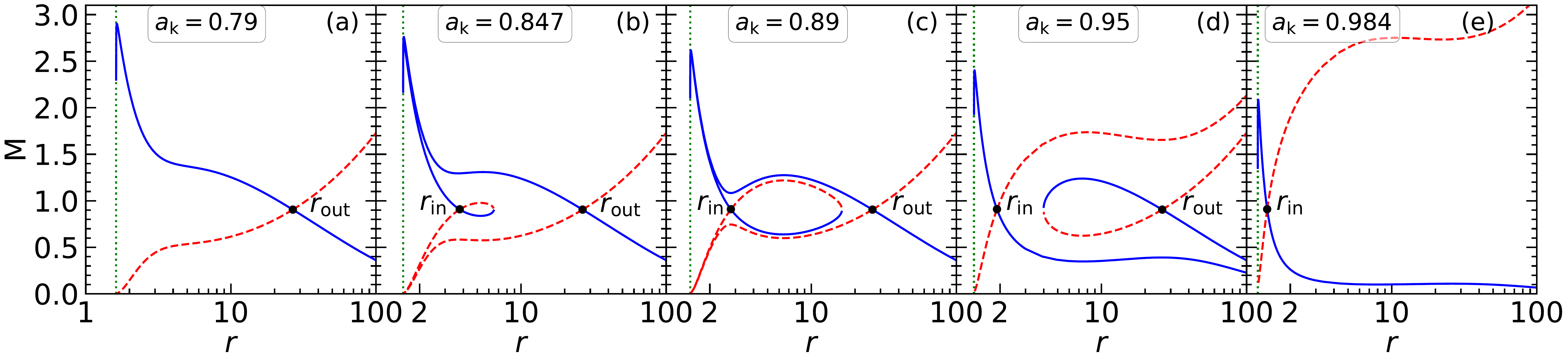}
    \end{center}
    \caption{Modification of transonic accretion solutions ($M$ vs $r$) with the increase of $a_{\rm k}$ as marked in each panel. Here, we fix the model parameters as $\mathcal{E} = 1.0084$, $\lambda = 2.1$, and $\beta=0.05$, respectively. Solid (blue) curves denote accretion solutions, whereas dashed (red) curves are for winds. Filled circles refer critical points ($r_{\rm in}$ and/or $r_{\rm out}$). Dotted vertical line denotes the throat radius as (a) $r_{\rm th}=1.6196$, (b) $r_{\rm th}=1.5387$, (c) $r_{\rm th}=1.4639$, (d) $r_{\rm th}=1.3226$ and (e) $r_{\rm th}=1.1942$. See the text for the details.
    }
    \label{mach_ak}
\end{figure*}

\subsection{Classification of Global Transonic Solutions}

Indeed, the nature of the transonic accretion solutions depends on the energy ($\cal E$) and angular momentum ($\lambda$) of the flow around WH. Towards this, in Fig. \ref{para-sol}a, we separate the effective domain of the parameter space in $\lambda-{\cal E}$ plane according to the nature of the transonic accretion solutions around WH. Here, we choose $a_{\rm k} = 0.99$ and $\beta = 0.05$, and identify four distinct regions in the parameter space that provide O-type, A-type, W-type  and I-type transonic accretion solutions. For the purpose of representation, we depict the typical examples of transonic accretion solutions from these four regions in panels (b-e) of Fig. \ref{para-sol}, where $M$ is plotted as function of $r$. These solutions are obtained for different sets of ($\lambda, {\cal E}$) chosen from the marked regions of the $\lambda-{\cal E}$ parameter space. In each panels, solid (blue) and dashed (red) curves represent flow solutions corresponding to accretion and wind, and filled circles denote the critical points ($r_{\rm in}$ and/or $r_{\rm out}$). In panel (b), we present the O-type solution which are obtained for ($\lambda, \mathcal{E})=(1.70,1.005$) and the solution possesses outer critical point at $r_{\rm out}=52.8253$ before advancing towards the WH throat ($r_{\rm th}$). We calculate A-type solution for ($\lambda, \mathcal{E})=(1.95,1.005$) and the obtained results are shown in panel (c). The solution of this kind contains both inner and outer critical points, and we find $r_{\rm in}=1.5384$ and $r_{\rm out}=50.0378$. The entropy accretion rate at $r_{\rm in}$ and $r_{\rm out}$ are computed as $\dot{\cal M}(r_{\rm in}) \equiv \dot{\cal M}_{\rm in}= 8.66 \times 10^7$ and $\dot{\cal M}(r_{\rm out}) \equiv \dot{\cal M}_{\rm out}= 7.337 \times 10^7$, respectively. Note that accretion solution passing through $r_{\rm out}$ successfully connects the outer edge of the disk ($r_{\rm edge}$) and the WH throat ($r_{\rm th}$), where solution containing $r_{\rm in}$ fails to do so. Next, we obtain W-type solution for ($\lambda, \mathcal{E})=(1.96,1.01$) that yields $r_{\rm in}=1.4995$ and $r_{\rm out}=23.1166$ (see panel (d)). We find that for W-type solutions, entropy accretion rate at $r_{\rm out}$ is higher than the entropy accretion rate at $r_{\rm in}$ as $\dot{\cal M}_{\rm out} = 11.474 \times 10^7$ and $\dot{\cal M}_{\rm in} = 8.2 \times 10^7$. We also notice that accretion solution possessing $r_{\rm out}$ can not extend up to the WH throat ($t_{\rm th}$), however, it seamlessly connects $r_{\rm edge}$ and $r_{\rm th}$ when passing through $r_{\rm in}$. Finally, the results corresponding to I-type solution is shown in panel (e) which possesses only inner critical point ($r_{\rm in}$), and results are obtained for ($\lambda, \mathcal{E})=(1.9,1.02$) with $r_{\rm in}=1.7160$.

\begin{figure}
    \begin{center}
    \includegraphics[width=\columnwidth]{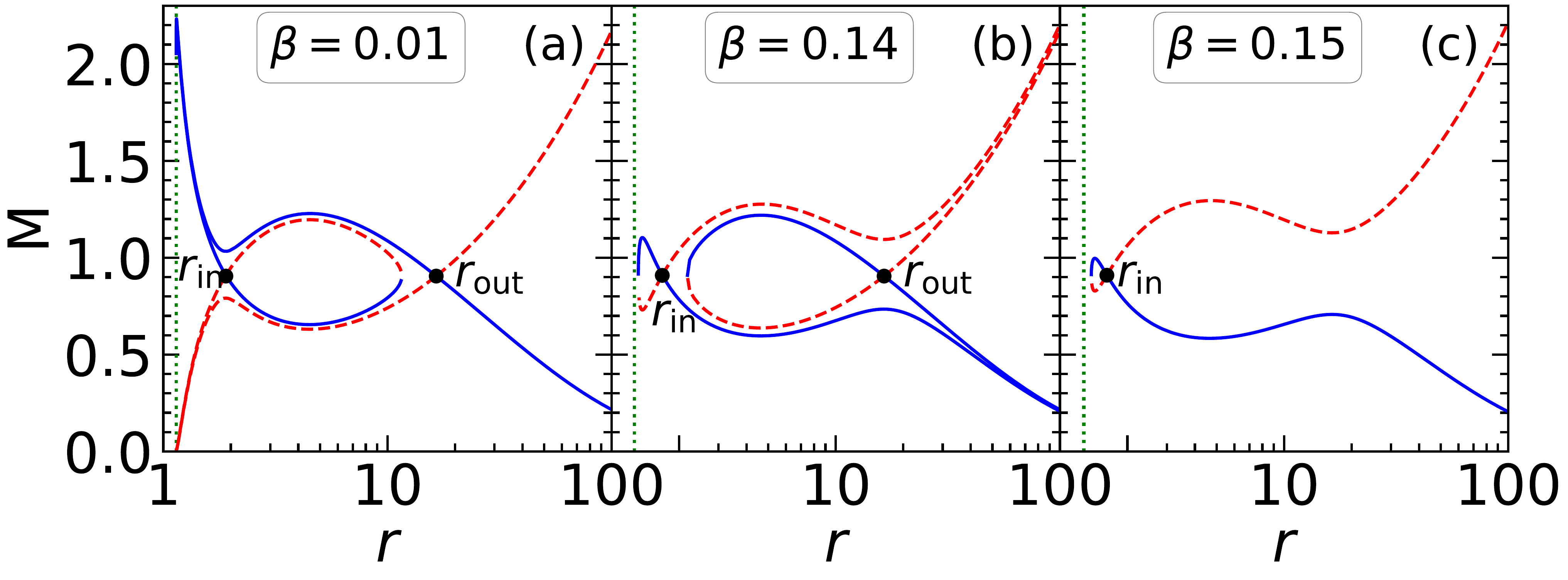}
    \end{center}
    \caption{Same as Fig. \ref{mach_ak}, but for different $\beta$ as marked in each panel. Here, we fix the model parameters as $\mathcal{E} = 1.0137$, $\lambda = 1.881$, and $a_{\rm k}=0.99$, respectively. Dotted vertical line denotes the throat radius as (a) $r_{\rm th}=1.1418$, (b) $r_{\rm th}=1.2634$ and (c) $r_{\rm th}=1.2782$, respectively. See the text for the details.
    }
    \label{mach_beta}
\end{figure}

\subsection{Modification of Global Transonic Solutions}

It is intriguing to examine the role of $a_{\rm k}$ in deciding the nature of the accretion solution around WH. In order for that we fix the model parameters as $\mathcal{E} = 1.0084$, $\lambda = 2.1$, and $\beta=0.05$ and calculate the flow solutions by tuning $a_{\rm k}$. The obtained results are depicted in Fig. \ref{mach_ak}, where solid curve denotes accretion solution and dashed curve is for wind. In panel (a), we obtain O-type solution for $a_{\rm k}=0.79$ having outer critical point at $r_{\rm out}=26.7261$ and throat radius at $r_{\rm th}=1.619$. When $a_{\rm k}$ is increased as $0.847$, we find that inner critical point appears at $r_{\rm in}=3.7780$ along with the outer critical point at $r_{\rm out}=26.5471$, as shown in panel (b). For $a_{\rm k}=0.89$, flow continue to possess multiple critical points at $r_{\rm in}=2.7921$ and $r_{\rm out}=26.4077$ (see panel c) and the overall character of the solution remains qualitatively same as in panel (b). As mentioned earlier that for accretion solutions of this kind, ${\dot {\cal M}}_{\rm in} > {\dot {\cal M}}_{\rm out}$. When $a_{\rm k}$ is increased further as $0.95$, the character of the solution alters, although it continues to possess multiple critical points at $r_{\rm in}=1.9050$ and $r_{\rm out}=26.2056$ (see panel d). For accretion solutions similar to this, we obtain ${\dot {\cal M}}_{\rm out} > {\dot {\cal M}}_{\rm in}$. Beyond a critical limit, such as $a_{\rm k}=0.984$, we notice that the outer critical point disappears and the flow solution passed through the inner critical point only at $r_{\rm in} = 1.3860$, as depicted in panel (e).

For the purpose of completeness, we examine the effect of $\beta$ in deriving the flow solutions. Towards this, we choose the model parameters as ${\cal E}=1.0137$, $\lambda=1.881$ and $a_{\rm k}=0.99$, and vary $\beta$ to compute the solutions. In Fig. \ref{mach_beta}, we present the obtained results where $\beta$ is increased in succession. We observe that $\beta=0.01$ provides A-type solution possessing multiple critical points at $r_{\rm in}=1.9016$ and $r_{\rm out}=16.5016$, as shown in Fig. \ref{mach_beta}a. As $\beta$ is increased to $0.14$, the nature of the accretion solution alters to W-type with $r_{\rm in} = 1.6829$ and $r_{\rm out}=16.4300$ (see Fig. \ref{mach_beta}b). For $\beta=0.15$, the outer critical point disappears and we obtain solution containing only inner critical point at $r_{\rm in}=1.6202$.

\subsection{Subsonic Accretion Solutions}

\begin{figure}
    \begin{center}
    \includegraphics[width=\columnwidth]{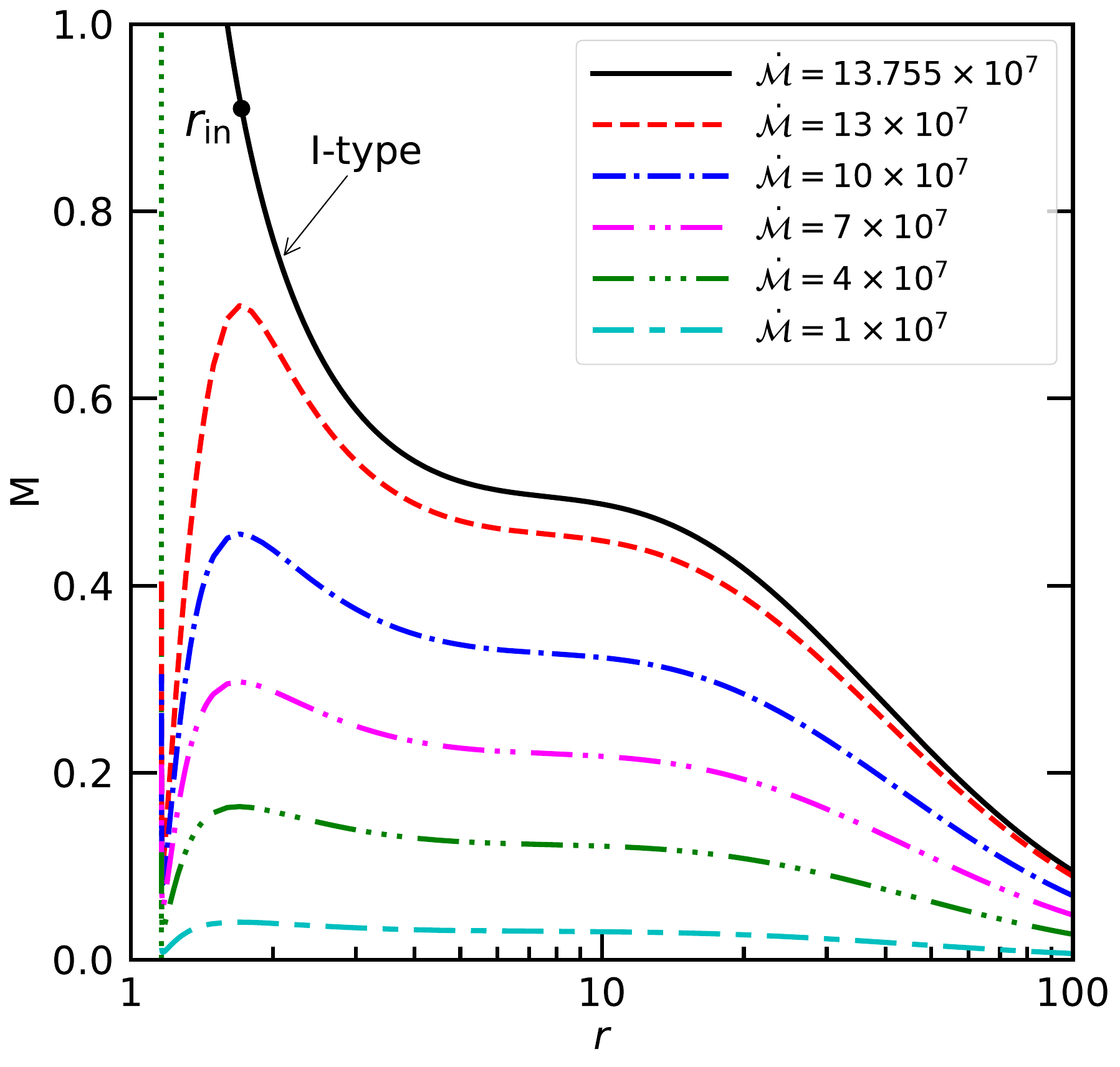}
    \end{center}
    \caption{Variation of Mach number ($M$) as a function of radial coordinate ($r$) for subsonic solutions associated with I-type transonic accretion solution. Here, we choose the model parameters as ${\cal E}=1.02$, $\lambda = 1.90$, $a_{\rm k}=0.99$ and $\beta=0.05$. Dashed (red), dot-dashed (blue), dot-dot-dashed (magenta), dot-dot-dot-dashed (green) and small-big-dashed (cyan) curves are for $\dot{\cal M} = 13 \times 10^7$, $10\times 10^7$, $7\times 10^7$, $4\times 10^7$, and $1\times 10^7$, respectively. Solid (black) curve refers the I-type transonic accretion solution (see Fig. \ref{para-sol}e) possessing entropy accretion rate as $\dot{\cal M}=13.755 \times 10^7$. Dotted vertical line denotes the throat radius $r_{\rm th}=1.1603$. See the text for the details.
 	}
    \label{sol_subs_i}
\end{figure}

\begin{figure}
    \begin{center}
    \includegraphics[width=\columnwidth]{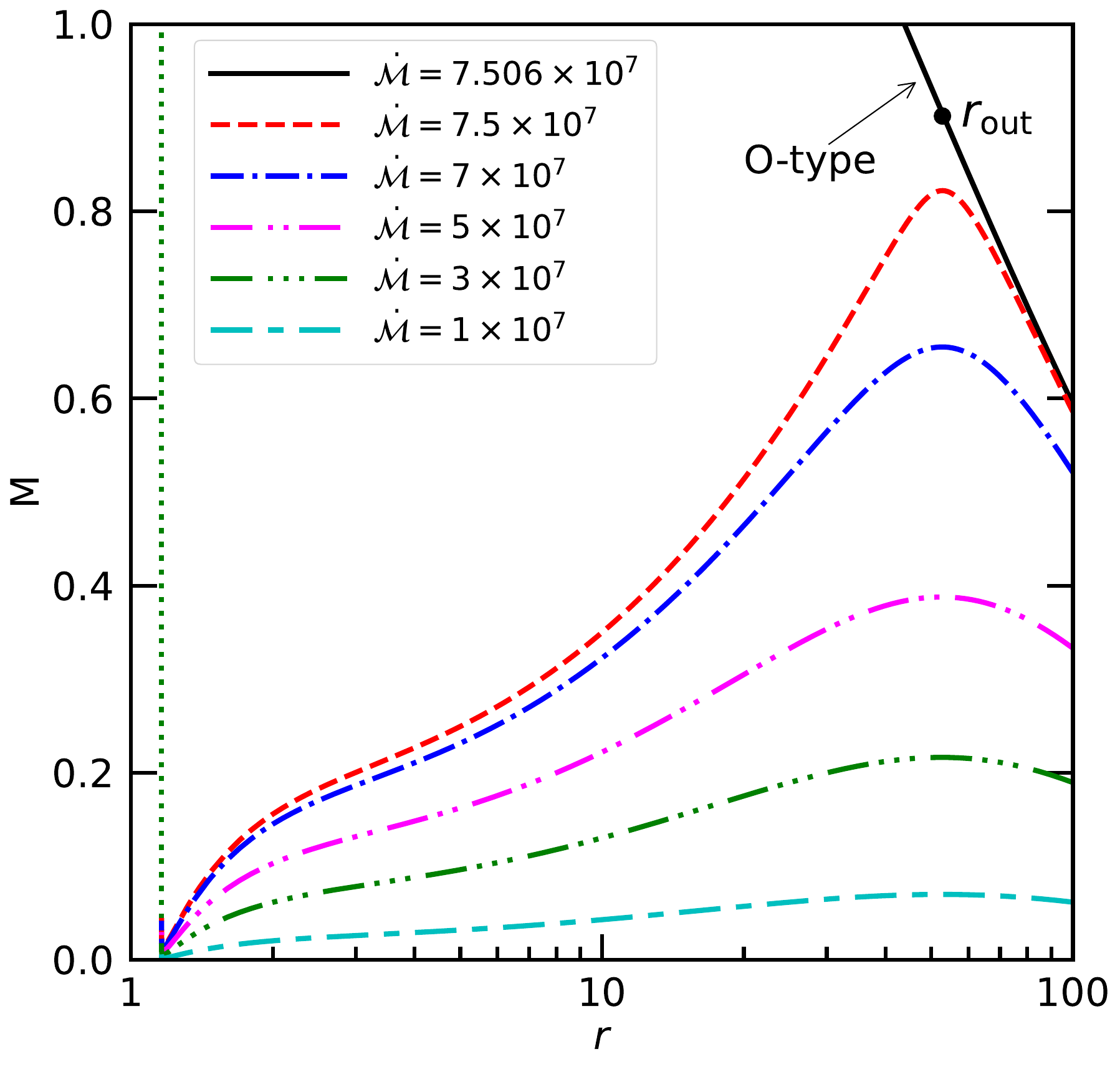}
    \end{center}
    \caption{Same as Fig. \ref{sol_subs_i}, but for subsonic solution associated with O-type transonic accretion solution (see Fig. \ref{para-sol}b). Here, we choose ${\cal E}=1.70$, $\lambda = 1.005$, $a_{\rm k}=0.99$ and $\beta=0.05$. Dashed (red), dot-dashed (blue), dot-dot-dashed (magenta), dot-dot-dot-dashed (green) and small-big-dashed (cyan) curves are for $\dot{\cal M} = 7.5 \times 10^7$, $7\times 10^7$, $5\times 10^7$, $3\times 10^7$, and $1\times 10^7$, respectively. Solid (black) curve refers the O-type transonic accretion solution with $\dot{\cal M}=7.505 \times 10^7$. Dotted vertical line denotes the throat radius $r_{\rm th}=1.1603$. See the text for the details.
 	}
    \label{sol_subs_o}
\end{figure}

As already mentioned, besides the transonic solutions, subsonic solutions are also exist around WH. Accordingly, we examine the nature of the subsonic solutions for flows with fixed model parameter (${\cal E}, \lambda, a_{\rm k}, \beta$). For this, we begin with a reference I-type transonic accretion solution obtained for ${\cal E}=1.02$, $\lambda = 1.90$, $a_{\rm k}=0.99$ and $\beta=0.05$ (see Fig. \ref{para-sol}e). The entropy accretion rate for this solution is computed as $\dot{\cal M} = 13.755 \times 10^7$. Now, we follow the method described in \S IV B to calculate the subsonic accretion solution around WH by decreasing $\dot{\cal M}$ while keeping all the remaining model parameters unchanged. The obtained results are depicted in Fig. \ref{sol_subs_i}, where the results presented with dashed (red), dot-dashed (blue), dot-dot-dashed (magenta), dot-dot-dot-dashed (green) and small-big-dashed (cyan) curves are obtained for $\dot{\cal M} = 13 \times 10^7$, $10\times 10^7$, $7\times 10^7$, $4\times 10^7$, and $1\times 10^7$, respectively. Interestingly, we observe that for a given set of model parameters (${\cal E},\lambda, a_{\rm k}, \beta$), $\dot{\cal M}$ always remains lower for subsonic solutions compared to the transonic solution (solid curve in black), which predominantly indicates that transonic solutions are thermodynamically preferred over the subsonic solutions because of their high entropy content. Further, in Fig. \ref{sol_subs_o}, we present the subsonic solution associated with the O-type transonic accretion solution. Here, the results are obtained by varying the entropy accretion rate as $\dot{\cal M} = 7.5 \times 10^7$, $7\times 10^7$, $5\times 10^7$, $3\times 10^7$, and $1\times 10^7$, keeping other model parameters fixed as ${\cal E}=1.70$, $\lambda = 1.005$, $a_{\rm k}=0.99$ and $\beta=0.05$. As in Fig. \ref{sol_subs_i}, here also we observe that $\dot{\cal M}$ is lower for subsonic solutions compared to the transonic accretion solution (solid curve in black) suggesting that transonic accretion solution are preferred over the subsonic solutions.

\subsection{Modification of Parameter Space for Multiple Critical Points}

\begin{figure}
    \begin{center}
    \includegraphics[width=\columnwidth]{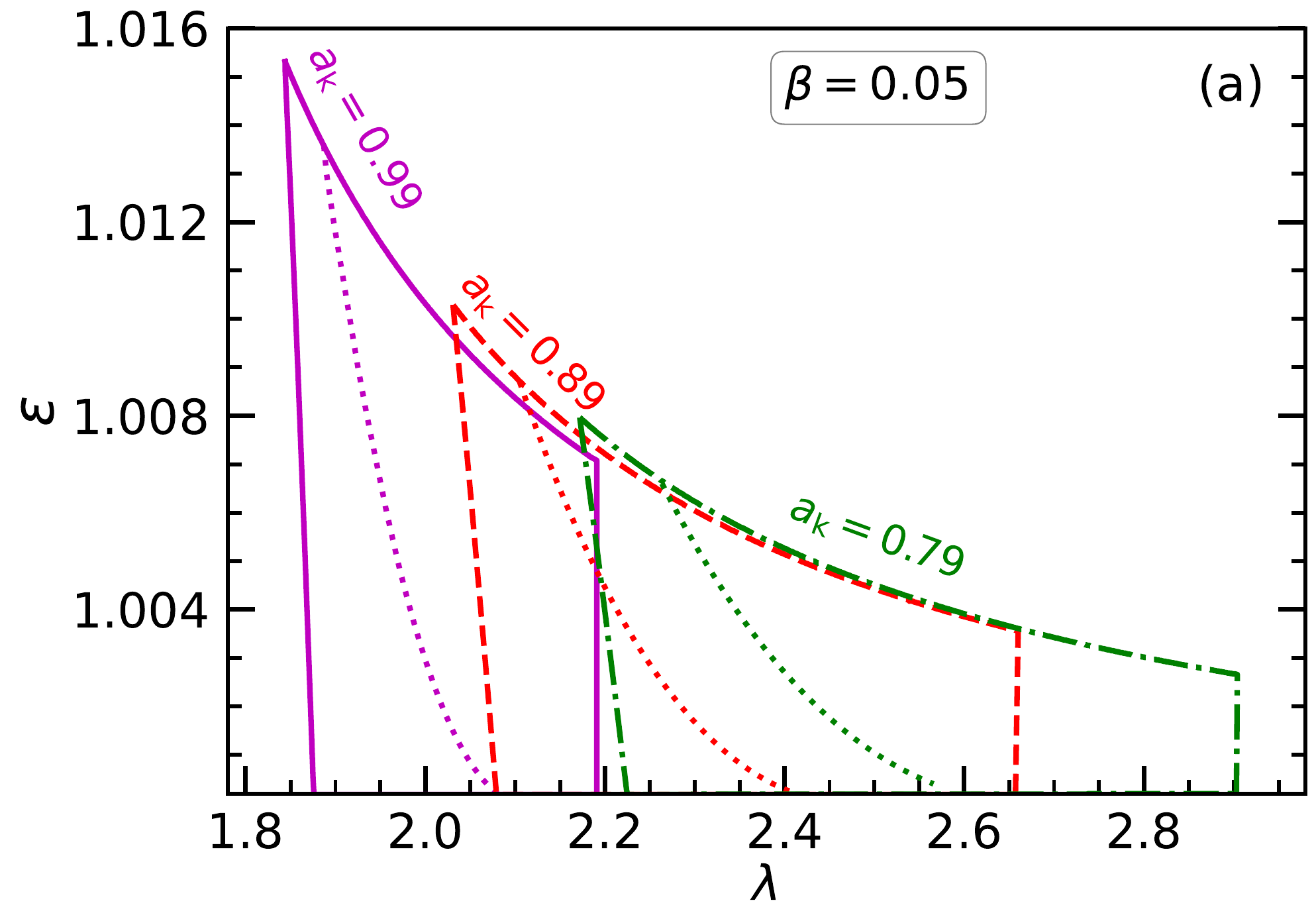}
    \includegraphics[width=\columnwidth]{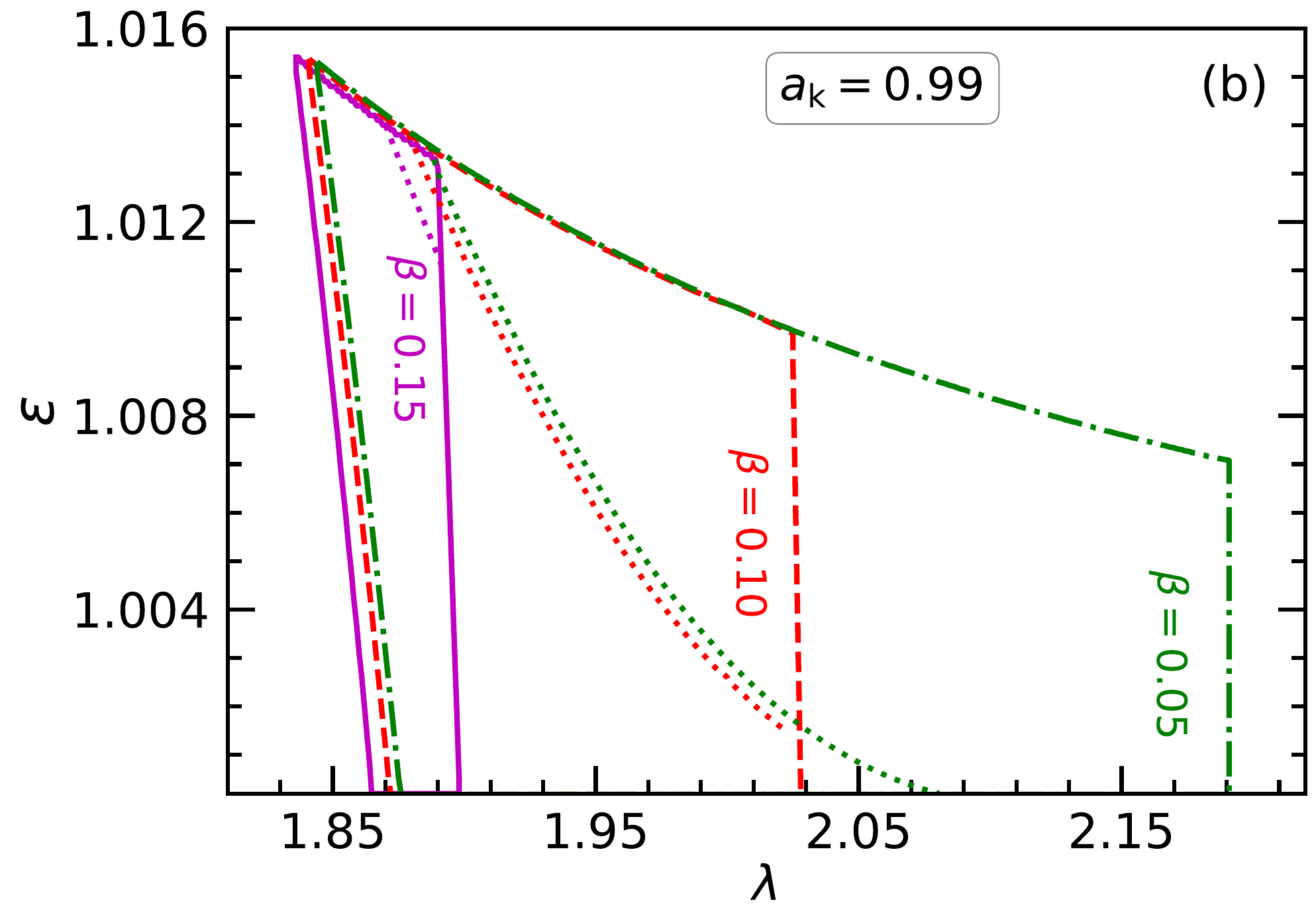}
    \end{center}
    \caption{Parameter space for multiple critical points in $\lambda-\mathcal{E}$ plane for different (a) $a_{\rm k}$ and (b) $\beta$ values. In panel (a), we choose $\beta = 0.05$ and the regions bounded with solid (magenta), dashed (red) and dot-dashed (green) curves are obtained for $a_{\rm k}= 0.99$, $0.89$ and $0.79$, respectively. Similarly, in panel (b), we fix $a_{\rm k}=0.99$, and solid (magenta), dashed (red) and dot-dashed (green) curves separate the region for $\beta = 0.15$, $0.10$ and $0.05$, respectively. Dotted curve separates the A-type and W-type solutions in each parameter space. See the text for the details.}
    \label{ds_para_en_lm_ak}
\end{figure}

It is noteworthy that depending on the model parameters, transonic flow possesses either single or multiple critical points. Following this, we identify the ranges of $\lambda$ and $\cal E$ that render multiple critical points while keeping $a_{\rm k}$ and $\beta$ fixed (see Fig. \ref{para-sol}). However, it is useful to examine the modification of $\lambda-{\cal E}$ parameter space due to the change of $a_{\rm k}$ and $\beta$ values. Towards this, in Fig. \ref{ds_para_en_lm_ak}a, we present how the effective domain of the parameter space alters due to the increase of $a_{\rm k}$ for a fixed $\beta$ value as $0.05$. Regions bounded with solid (magenta), dashed (red) and dot-dashed (green) curves are obtained for $a_{\rm k}=0.79$, $0.89$ and $0.99$, respectively. Each parameter space is further subdivided using dotted curve that separates A-type solutions (left side) from the W-type solutions (right side). We also observe that flow continues to possess multiple critical points for higher $a_{\rm k}$, provided $\lambda$ is relatively lower. This is expected because of the fact that the marginally stable angular momentum generally decreases with the increase of $a_{\rm k}$ due to the spin-orbit coupling embedded in the spacetime \cite{Das-Chakrabarti2008}. Similarly, in Fig. \ref{ds_para_en_lm_ak}b, we present the variation of the parameter space for different $\beta$. Here, we fix $a_{\rm k} = 0.99$, and boundaries drawn with dot-dashed (green), dashed (red) and solid (magenta) curves separate the regions for $\beta=0.05$, $0.10$ and $0.15$, respectively. We observe that for a fixed $a_{\rm k}$, the effective domain of the parameter space is shrunk with the increase of deformation parameter $\beta$.
 
\begin{figure}
    \begin{center}
    \includegraphics[width=\columnwidth]{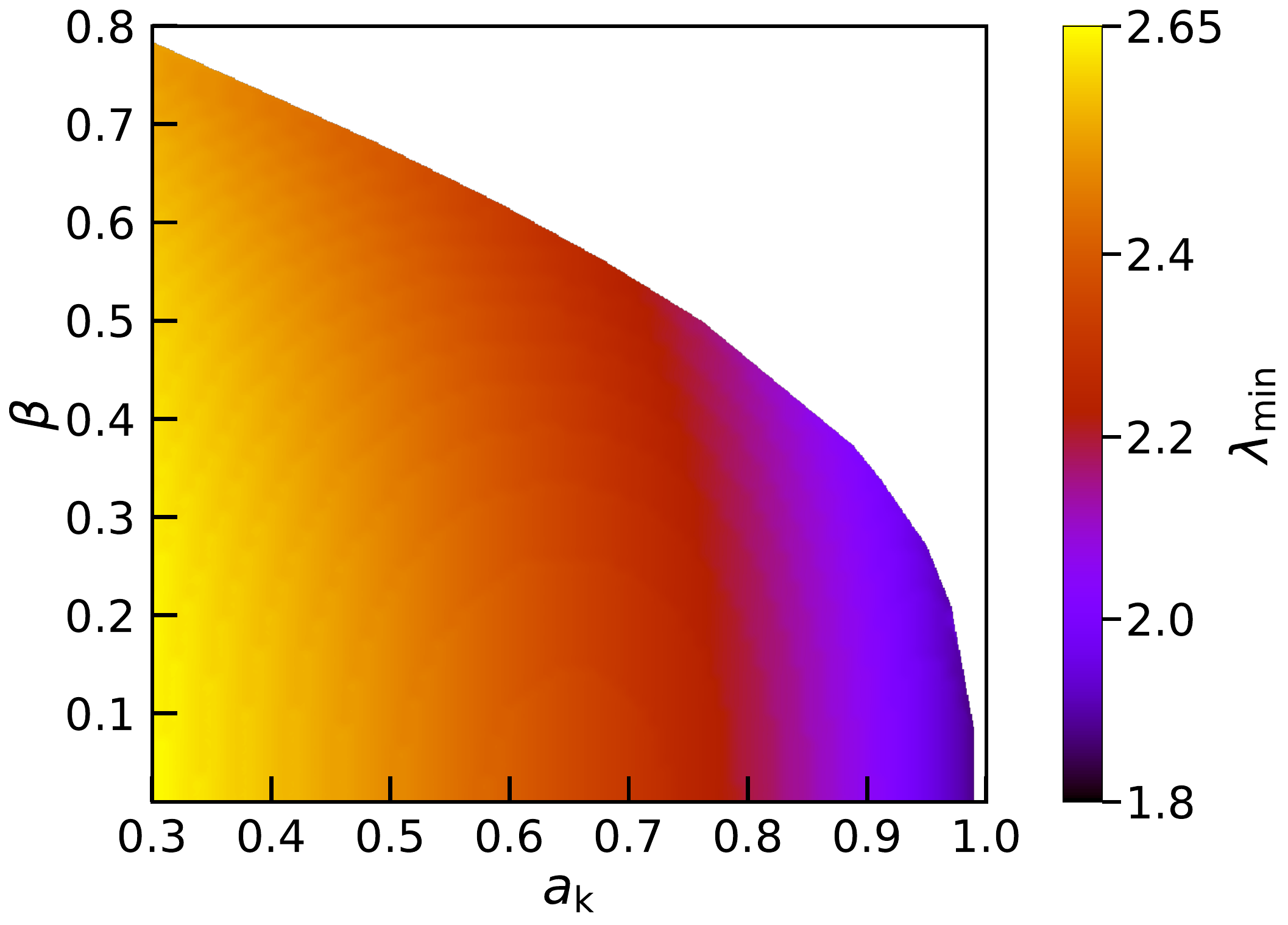}
    \end{center}
    \caption{Two-dimensional projection of the three-dimensional plot of $a_{\rm k}$, $\beta$ and $\lambda_{\rm min}$ for solutions containing multiple critical points. Here, we choose energy $\mathcal{E} = 1.004$. The colorbar represents the range of minimum angular momentum ($\lambda_{\rm min}$). See the text for the details.}
    \label{ds_para_beta_ak_fixed_en}
\end{figure}

Moreover, it is compelling to analyze the range of $\beta$ that renders multiple critical points as well. In doing so,  for the purpose of representation, we fix the energy of the flow as ${\cal E}=1.004$, and freely vary angular momentum ($\lambda$) to find its minimum value ($\lambda_{\rm min}$) yielding multiple critical points for $\beta \ge 0$ and $0 \le a_{\rm k} < 1$. Here, we focus on $\lambda_{\rm min}$ as it coarsely interprets the limiting value describing the quasi-radial nature of the flow containing multiple critical points. The obtained results are plotted in Fig. \ref{ds_para_beta_ak_fixed_en}, where two-dimensional projection of the three-dimensional plot spanned with $a_{\rm k}$, $\beta$ and $\lambda_{\rm min}$. In the figure, vertical colorbar denotes the range as $1.8 \le \lambda_{\rm min} \le 2.65$. Figure evidently indicates that the range of $\beta$ is decreased with the increase of $a_{\rm k}$, and $\lambda_{\rm min}$ anti-correlates with $a_{\rm k}$. 

\begin{figure}
    \begin{center}
    \includegraphics[width=\columnwidth]{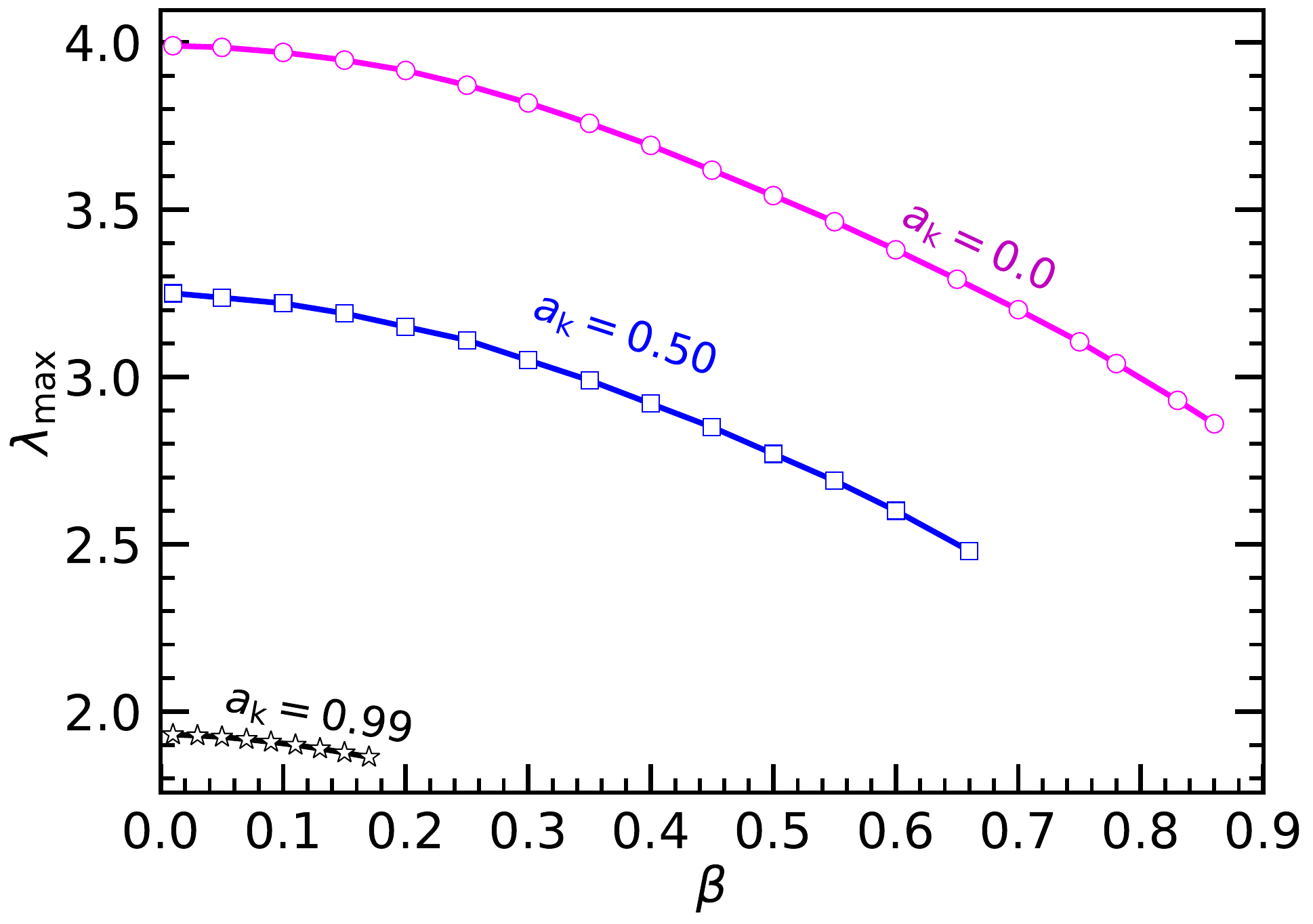}
    \end{center}
    \caption{Variation of $\lambda_{\rm max}$ with $\beta$ for three different values of $\a_{\rm k}$ yielding solutions possessing multiple critical points. Open circles, squares and asterisks joined with solid lines represent results corresponding to $a_{\rm k}= 0.0, 0.50,$ and $0.99$, respectively. See the text for the details.}
    \label{ds_para_beta_lm_max}
\end{figure}

Next, we put effort to calculate the upper limit of angular momentum ($\lambda_{\rm max}$) that renders multiple critical points. The obtained results are presented in Fig. \ref{ds_para_beta_lm_max}, where we illustrate the variation of $\lambda_{\rm max}$ as function of $\beta$ for different $a_{\rm k}$ values. Open circles, squares and asterisks joined with solid lines denote the results obtained for $a_{\rm k}=0.0$, $0.5$ and $0.99$, respectively. Here, energy of the flow is varied freely. We observe that for a fixed $a_{\rm k}$, $\lambda_{\rm max}$ monotonically decreases with the increase of $\beta$, and as $a_{\rm k}$ is increased, the allowed range of $\beta$ for multiple critical points decreases. We also notice that for a given $\beta$, when $a_{\rm k}$ is higher, $\lambda_{\rm max}$ becomes lower and vice versa.

\section{Radiative Emission Properties}

In this section, we examine the disk luminosity ($L$) focusing on free-free emission as it is regarded as one of the relevant radiative mechanism active inside the convergent single temperature accretion flow \cite[][and references therein]{Sarkar-Das2016,Okuda-etal2019}. Accordingly, we calculate $L$ as,
\begin{equation}
    L=2 \int_{0}^{\infty}\int_{r_{\rm th}}^{r_{\rm edge}} \int^{2\pi}_{0}(Hr)\epsilon(\nu_e)d\nu_o dr d\phi.
    \label{luminosity}
\end{equation}
Here, $\epsilon(\nu)$ denotes the bremsstrahlung emissivity at frequency $\nu$ and is given by \cite{Vietri-2008},
\begin{equation}
    \epsilon(\nu)=\frac{32\pi e^6}{3m_e c^3}\left(\frac{2\pi}{3k_Bm_eT_e}\right)^{1/2} Z_i^2 n_e n_i e^{-h\nu/k_BT_e}g_{br},
    \label{emissivity}
\end{equation}
where $m_e$ and $e$ are mass and charge of the electron, $k_B$ is the Boltzmann constant, $h$ is the Planck’s constant, $\nu$ is the frequency, $Z_i$ is the ion charge, and $g_{br}$ is the Gaunt factor \cite{Karzas-1961} assumed to be unity. In this work, we consider single temperature flow and following \cite{Chattopadhyay-Chakrabarti2002}, we estimate electron temperature as $T_e = \sqrt{(m_e/m_i)}T$,	where $T$ denotes the flow temperature and $m_i$ is the ion mass. The emitted frequency ($\nu_e$) is related to the observed frequency ($\nu_o$) as $\nu_e = (1 + z)\nu_o$, where $z$ denotes the red-shift factor. Following \cite{Luminet-1979}, we determine $z$ considering fixed inclination angle $i=\pi/4$ for Kerr-like WH. In addition, we choose $M_{\rm WH}=10M_\odot$ and ${\dot m} = 0.1$ while computing disk luminosity.

\begin{figure}
    \begin{center}
    \includegraphics[width=\columnwidth]{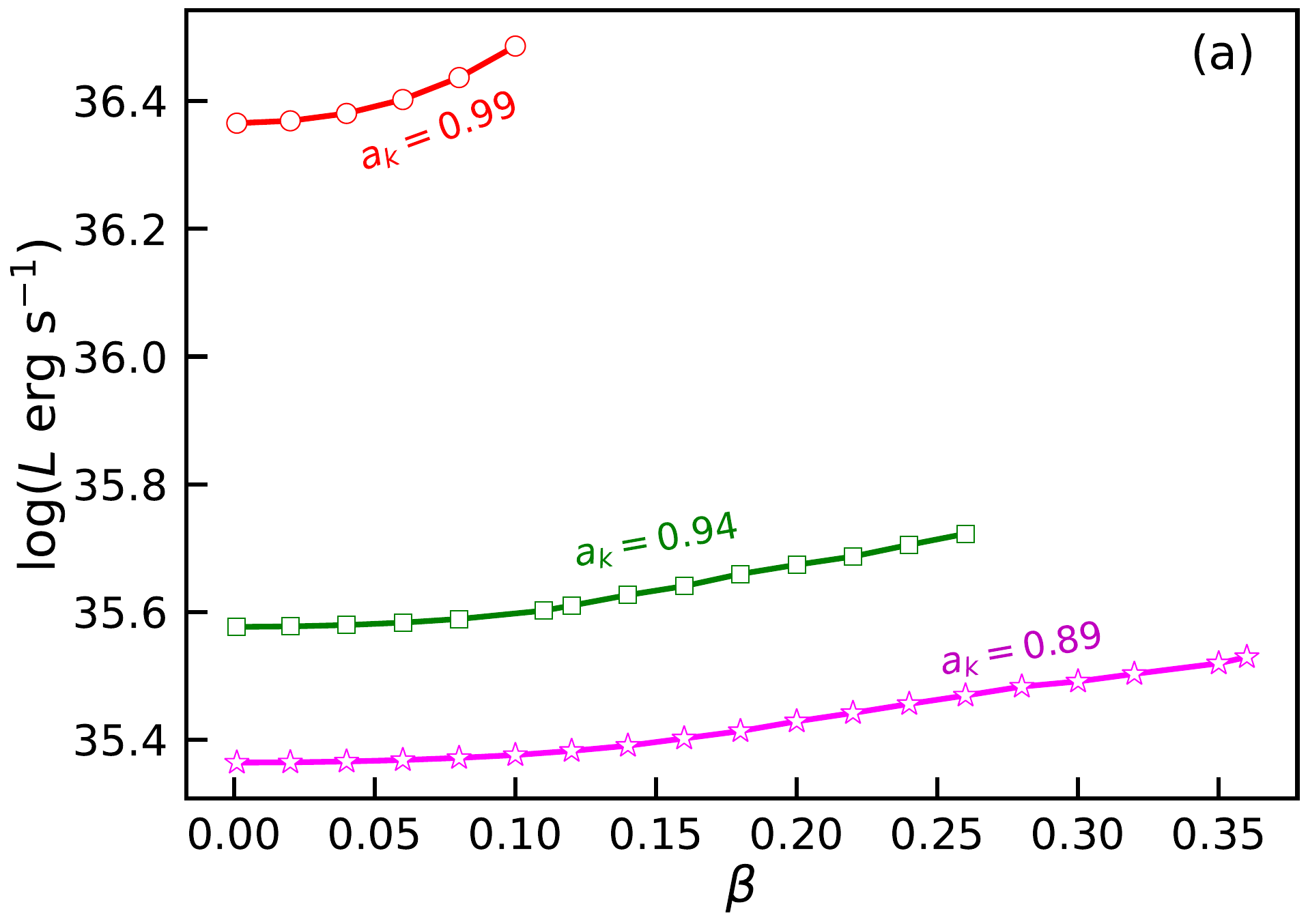}
    \includegraphics[width=\columnwidth]{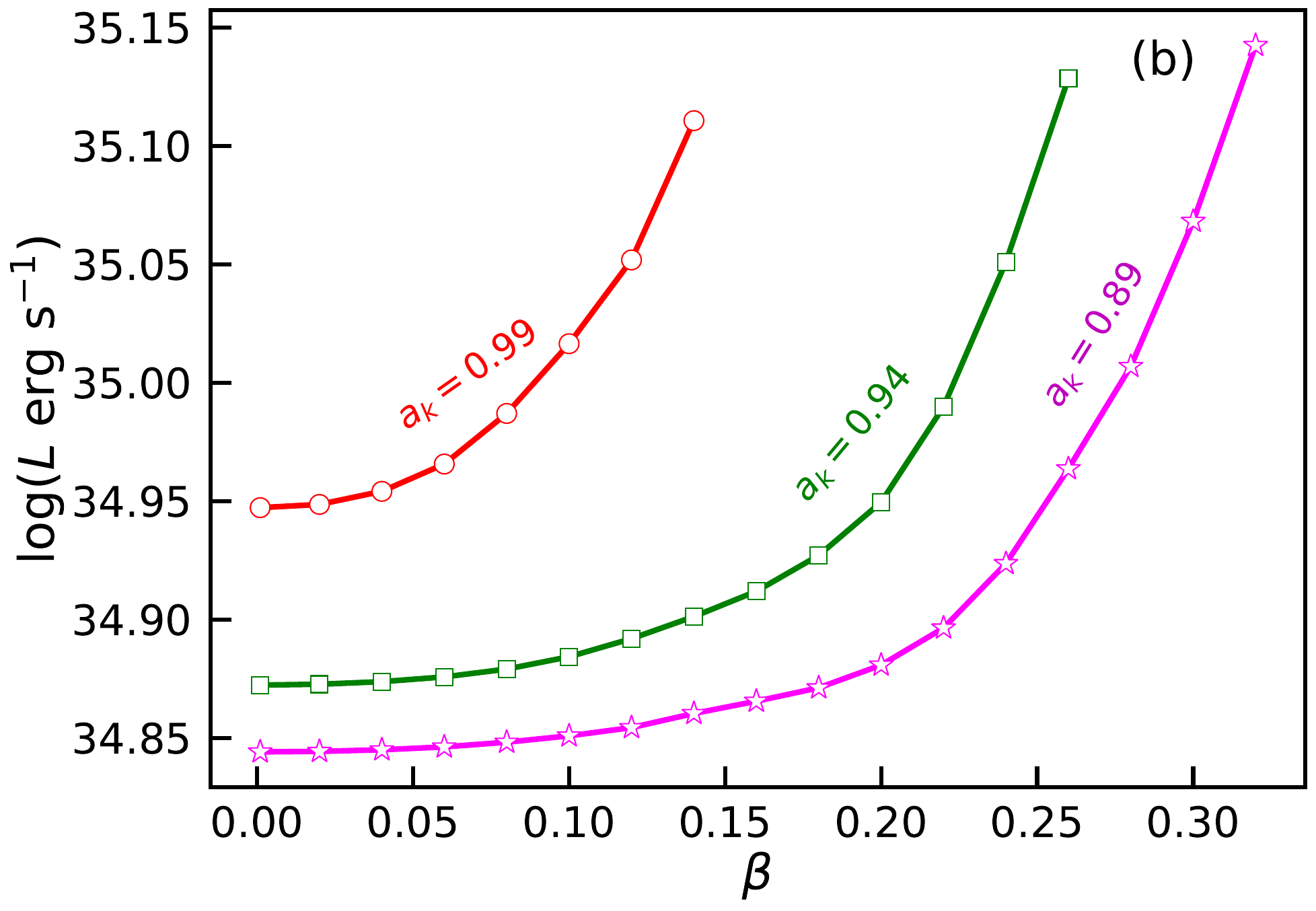}
    \end{center}
    \caption{Variation of disk luminosity ($L$) as function of $\beta$ for different $a_{\rm k}$. In panel (a), results corresponding to I-type accretion solutions are depicted for flow with $\mathcal{E}=1.02$ and $\lambda=1.90$. Open circles, squares and asterisks joined with solid lines are for $a_{\rm k}=0.99$, $0.94$, and $0.89$, respectively. In panel (b), results same as panel (a) are shown, but for O-type accretion solutions, where $\mathcal{E}=1.004$ and $\lambda=1.85$ are chosen. See the text for the details.}
    \label{tran_lumi_rin}
\end{figure}

We present the obtained results in Fig. \ref{tran_lumi_rin}, where the variation of disk luminosity ($L$) with $\beta$ for different $a_{\rm k}$ is depicted. The results corresponding to I-type and O-type solutions are presented in panel (a) and (b). In both panels, open circles, squares and asterisks joined with solid lines denote results for $a_{\rm k}=0.99$, $0.94$ and $0.89$, respectively. We find that for a fixed $a_{\rm k}$, $L$ increases with the increase of $\beta$. Similarly, when $\beta$ is kept fixed, $L$ is seen to increase for higher $a_{\rm k}$. Overall, we observe that for a fixed set of ($a_{\rm k},\beta$), I-type solutions yields higher disk luminosity compared to the same obtained from O-type solutions. This happens because I-type solutions exhibit higher density profile compared to the O-type solutions as I-type flow remains subsonic in the range $r_{\rm in} < r \le r_{\rm edge}$.

\begin{figure}
    \begin{center}
    \includegraphics[width=\columnwidth]{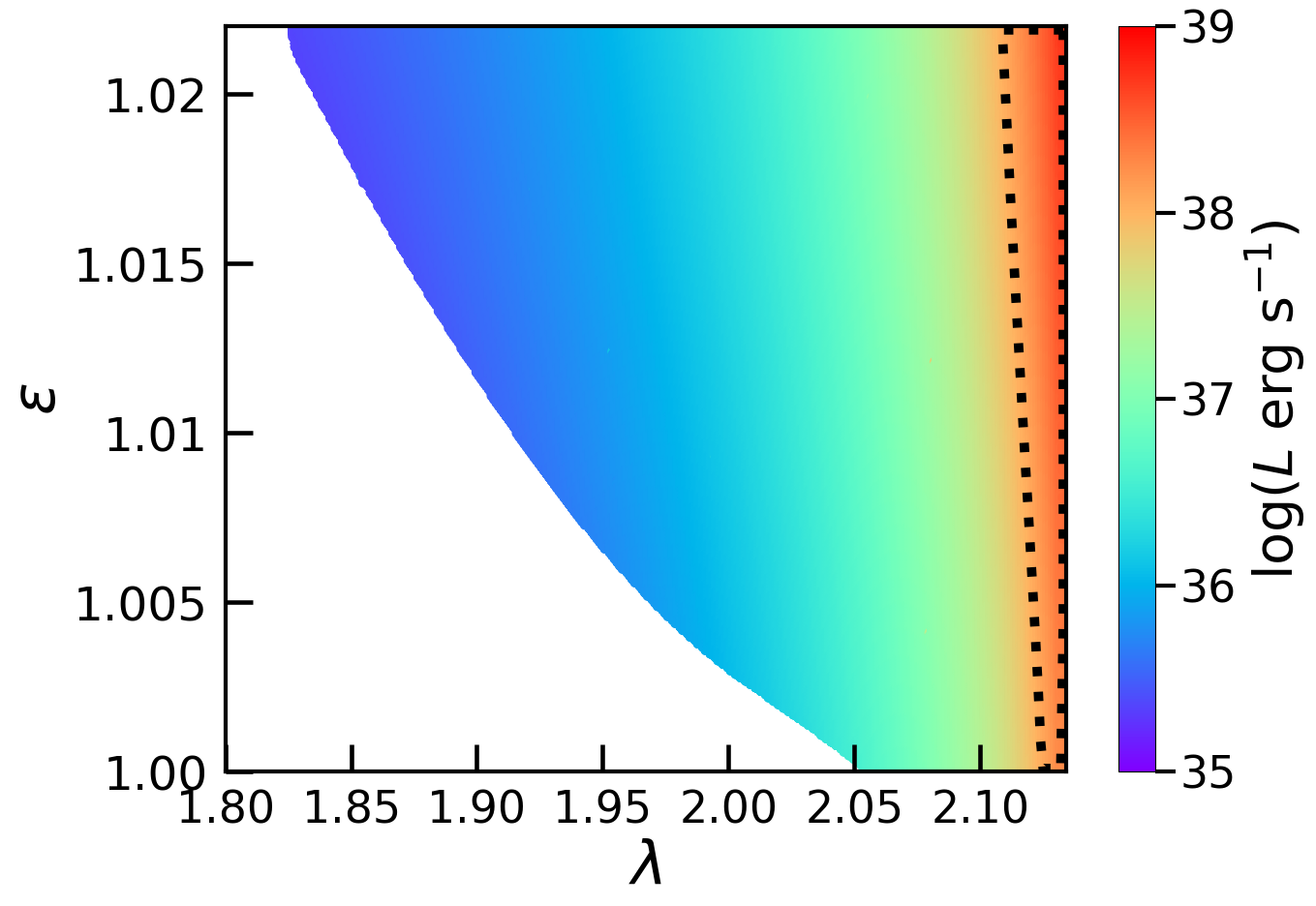}
    \end{center}
    \caption{Two-dimensional projection of the three-dimensional plot of $\mathcal{E}$, $\lambda$ and $\log(L$ $\rm erg~s^{-1})$ for transonic flow  due to free-free emission. The colorbar denotes the range of luminosity values. The region enclosed by the dotted curve yields disk luminosity consistent with the observed luminosity of Cygnus X$-3$ during its hypersoft state. See the text for the details.}
    \label{tran_cyg}
\end{figure}

We also put effort to explain the luminosity of a compact object Cygnus X-3, using our model formalism. Cygnus X-3 displays intense luminosity, predominantly in X-ray wavelengths. This sustained brightness, amidst its erratic behavior, hints at underlying mechanisms continuously fueling its emissions. Moreover, Cygnus X-3 exhibits a unique hypersoft state characterized by its bolometric X-ray flux reaching peak values in the range $2-8\times10^{-8}$ erg cm$^{-2}$ s$^{-1}$ \cite{Hjalmarsdotter-etal2008,Koljonen-etal2018}. Adopting the source distance of $7.4$ kpc \cite{McCollough-etal2016}, the source luminosity is estimated as $L_S \sim 1-5\times10^{38}$ erg s$^{-1}$ \cite{Zdziarski-etal2012}. In order to explain $L_S$, we compute the `model predicted' disk luminosity ($L$) arising from free-free emission for transonic accretion solutions around WH. In doing so, we use the source mass as $M_{\rm WH}= 2.4 M_{\odot}$ \cite{Zdziarski-etal2013}, and for the purpose of representation, we consider typical accretion rate $\Dot{m}=0.1$, source spin $a_{\rm k}=0.99$ and $\beta = 0.001$. The obtained results are presented in Fig. \ref{tran_cyg}, where we illustrate the two-dimensional projection of the three-dimensional plot of $\lambda$, $\mathcal{E}$ and $\log~(L$ erg s$^{-1})$. In the figure, the vertical colorbar denotes the disk luminosity in the range $35\le \log~(L~{\rm erg~s}^{-1}) \le 39$. In the figure, we identify a region bounded with dotted curves that yields the luminosity $1 \times 10^{38} \lesssim L \lesssim 5\times 10^{38}$ erg s$^{-1}$. These findings evidently indicate that our analysis in turn renders the representative values of the luminosity ($L_S$) of Cygnus X-3. Moreover, we argue that present model formalism seems to be potentially promising in explaining the luminosity of compact X-ray sources.

\section{Summery and  Conclusions}

In this work, we study the low angular momentum, inviscid, advective accretion flow around a stationary axisymmetric Kerr-like WH spacetime. The Kerr-like WH is characterized by the spin parameter $(a_{\rm k})$ and dimensionless parameter ($\beta$) along with its mass ($M_{\rm WH}$). In doing so, we examine the steady state accretion solutions which are obtained by solving the governing equations describing the accretion flow confined around the disk equatorial plane. Further, we investigate the role of $a_{\rm k}$ and $\beta$ in regulating the accretion dynamics. With this, we summarize our key findings in the below.

\begin{itemize}

    \item We calculate the transonic accretion solution (I-type) that passes through the inner critical point $r_{\rm in}$ around WH (see Fig. \ref{sol_i}). We find that the radial profile of the flow variables corresponding to this solution, such as density ($\rho$), pressure ($p$) and temperature ($T$) follow power-law distributions as $\rho \propto r^{-(n+2/5)}$, $p \propto r^{-(n+1)}$ and $T \propto r^{-(n-1/3)}$ with $n \sim 1$ inside the disk (see Fig. \ref{sol_prop}). However, solution deviates from self-similarity close to $r_{\rm th}$ mainly due to the non-linearity present in the WH spacetime.

    \item Further, for the first time to the best of our knowledge, we obtain the complete set of transonic accretion solutions (O-type, A-type, W-type and I-type) around WHs by tuning the model parameters, namely energy ($\cal E$), angular momentum ($\lambda$), spin parameter ($a_{\rm k}$), and $\beta$. We find that a given type of accretion solutions are not isolated solutions as these solutions continue to exist for wide range of model parameters. We also separate the domains of the parameter space in $\lambda-{\cal E}$ plane according to the nature of the accretion solutions (see Fig. \ref{para-sol}). Furthermore, we investigate the impact of $a_{\rm k}$ ($\beta$) values in altering the parameter space for multiple critical points. Our findings reveal that when $a_{\rm k}$ ($\beta$) is increased keeping $\beta$ ($a_{\rm k}$) fixed, the parameter space shifts towards the higher energy and lower angular momentum domain (see Fig. \ref{ds_para_en_lm_ak}). 

    \item We examine the role of $a_{\rm k}$ and $\beta$ in obtaining the transonic accretion solutions. We observe that for fixed $\cal E$, $\lambda$, and $\beta$ ($a_{\rm k}$), accretion solution alters its character as $a_{\rm k}$ ($\beta$) is increased (see Fig. \ref{mach_ak} and Fig. \ref{mach_beta}). This findings evidently indicate that both $a_{\rm k}$ and $\beta$ play pivotal role in deciding the nature of the transonic accretion solutions around WH.

    \item We further emphasize that subsonic accretion solutions are also possible around WH (see Fig. \ref{sol_subs_i}-\ref{sol_subs_o}). However, for fixed $\cal E$, $\lambda$, $a_{\rm k}$ and $\beta$, these solutions possess lower entropy content compared to the transonic solutions. Hence, we argue that transonic solutions around WH are thermodynamically preferred over the subsonic solutions.

    \item We compute the disk luminosity ($L$) considering bremsstrahlung emission and observe strong dependency of $L$ on both $a_{\rm k}$ and $\beta$. It becomes evident that for a fixed $\beta$ ($a_{\rm k}$), increasing $a_{\rm k}$ ($\beta$) leads to higher $L$ for both I-type and O-type transonic accretion solutions. In addition, we note that I-type solutions yield higher $L$ compared to O-type solutions (see Fig. \ref{tran_lumi_rin}). 
    
    \item We indicate that our model successfully elucidates the luminosity of compact X-ray source Cygnus X-3 during its hypersoft state. Based on this finding, we mention that the present model formalism offers the valuable insights of the accretion flow dynamics around WH that could drive the energetic emissions observed from enigmatic compact X-ray sources.
    
\end{itemize}

Finally, we state that this work is developed based on some assumptions. We neglect the effect of viscosity that usually takes care the angular momentum transport inside the disk allowing the matter to accrete towards the WH. We avoid magnetic fields although it is ubiquitous in all astrophysical sources. We also ignore the massloss from the disk which seems relevant in explaining disk-jet symbiosis commonly observed in Galactic black hole X-ray binaries. Indeed, implementation of these physical processes is beyond the scope of the present work. However, we intend to address them in our future endeavours.
 
\section*{Acknowledgement}
	
SD acknowledges financial support from Science and Engineering Research Board (SERB), India grant MTR/2020/000331.
	
\bibliographystyle{unsrtnat}
\bibliography{reference.bib}

\end{document}